\documentclass[aps,showpacs,superscriptaddress,nofootinbib]{revtex4}
\usepackage[dvips]{graphicx}
\usepackage{epsfig,rotating}
\usepackage{amsmath,amssymb}
\numberwithin{equation}{section}

\newcommand{\dbarq}{\frac{d^3q}{(2\pi)^3}}
\newcommand{\vx}{\vec{x}}
\newcommand{\vp}{\vec{p}}
\newcommand{\vq}{\vec{q}}
\newcommand{\vk}{\vec{k}}
\newcommand{\bw}{\overline{\omega}}
\newcommand{\bk}{\overline{k}}
\newcommand{\uvk}{\widehat{\bf{k}}}
\newcommand{\uvq}{\widehat{\bf{q}}}
\newcommand{\be}{\begin{equation}}
\newcommand{\ee}{\end{equation}}
\newcommand{\bea}{\begin{eqnarray}}
\newcommand{\eea}{\end{eqnarray}}
\begin{document}
\title{Neutrino collective excitations in the Standard Model at high temperature. }
\author{D. Boyanovsky}
\email{boyan@pitt.edu} \affiliation{Department of Physics and
Astronomy, University of Pittsburgh, Pittsburgh, Pennsylvania 15260,
USA}

\date{\today}

\begin{abstract}
Neutrino collective excitations are studied in the Standard Model at
high temperatures below the symmetry breaking scale in the regime
$T\gg M_{W,Z}(T)\gg gT$. Two parameters determine the properties of
the collective excitations: a  mass scale $m_\nu=gT/4$ which
determines the \emph{chirally symmetric} gaps in the spectrum and
$\Delta=M^2_W(T)/2m_\nu T$. The spectrum consists of left handed
negative helicity quasiparticles, left handed positive helicity
quasiholes and their respective antiparticles. For $\Delta <
\Delta_c = 1.275\cdots$ there are two gapped quasiparticle branches
and one gapless and two gapped quasihole branches, all but the
higher gapped quasiparticle branches terminate at end points. For
$\Delta_c < \Delta < \pi/2$ the quasiparticle spectrum features a
pitchfork bifurcation and for $\Delta >\pi/2$ the collective modes
are gapless quasiparticles with dispersion relation below the light
cone for $k\ll m_\nu$ approaching the free field limit for $k\gg
m_\nu$ with a rapid crossover between the soft non-perturbative to
the hard perturbative regimes for $k\sim m_\nu$.The \emph{decay} of
the vector bosons leads to a \emph{width} of the collective
excitations  of order $g^2$ which is explicitly obtained in the
limits $k =0$ and $k\gg m_\nu \Delta$. At high temperature this
damping rate is shown to be competitive with or larger than the
 collisional damping rate of order $G^2_F$ for a wide range of
neutrino energy.

\end{abstract}

 \maketitle

\section{Introduction}\label{sec:intro}

Neutrinos are emerging as  the bridge between particle physics,
astrophysics, cosmology and nuclear physics. Physical processes
involving neutrinos in dense and hot media are important from
cosmology\cite{book1,book2,book3,raffelt,pantaleone,dolgov,gouvea}
to the astrophysics of compact
stars\cite{prakash,reddy,yakovlev,barko}. More recently, it has been
proposed that leptogenesis in the early Universe could be the main
mechanism that explains the origin of the baryon
asymmetry\cite{fukugita,yanagida,buch}. In this scenario the
spectrum of neutrinos is  an important ingredient that determines
the size of the corrections in the  transport equations for
leptogenesis\cite{buch}. More refined studies of thermal
leptogenesis  include  finite temperature corrections in the fermion
propagators\cite{giudice}, which were previously found in
vector-like theories (QCD and QED)\cite{klimov,weldon}.

Neutrino propagation in hot and or dense matter has become very
important in astrophysics and cosmology after
Wolfenstein\cite{wolfenstein} studied the effective potential and
refractive index of neutrinos in matter and Mikheyev and Smirnov
recognized the possibility that matter effects result in  resonant
flavor oscillations\cite{smirnov}. The MSW effect is now recognized
to be the solution of the solar neutrino problem. The importance of
medium effects in the propagation of neutrinos warrants an active
program to study the possible cosmological and astrophysical
consequences of novel dispersion relations and non-equilibrium
aspects of neutrinos in  hot and dense media.

Early studies of neutrino propagation in hot and or dense media
focused on the possible modifications of the neutrino dispersion
relations in the regime of temperatures relevant for stellar
evolution or for big bang nucleosynthesis, namely up to $\sim 1-10
~\textrm{Mev}$. Neutrino dispersion relations and damping rates in
this regime of temperature has originally been obtained in
ref.\cite{notzold} up to lowest order in the momentum dependence of
the weak interactions in the standard model at temperatures well
below the electroweak scale. Since then these results have been
extended by several authors to include the corrections from light
scalars\cite{nieves}, to study matter effects in the oscillations of
neutrinos in the early
universe\cite{barbieri,enqvist,dolgov,dolivoOS} and to a medium that
includes nucleons, leptons and neutrinos\cite{dolivoDR} as well as
the relaxation (damping) rate of neutrinos in a hot and dense
medium\cite{notzold,tututi,thomson,barbieri,dolgov}. Damping rates
and mean free paths of neutrinos in models beyond the standard model
have been studied in ref.\cite{vilja}.

While all of these thorough studies of the aspects of neutrino
propagation focus on temperature (and chemical potential) scales
suitable for primordial nucleosynthesis or stellar evolution, a
systematic study of neutrino dispersion relations at high
temperatures, near the \emph{electroweak scale} has not (to the best
of our knowledge) yet emerged.

Motivated to a large extent to explore the possibility of thermal
leptogenesis in the standard model, as well as to study in general
possible non-equilibrium aspects of neutrinos that could be
relevant to lepto or baryogenesis\cite{kaplan}  the purpose of
this article is precisely to bridge this gap and study the
propagation of neutrinos at high temperatures, near (but below)
the electroweak scale \emph{in the standard model}.

Since in the standard model baryon number and CP are not conserved
it was conjectured that the cosmological baryon asymmetry could have
been generated at the electroweak phase transition, if it was of
first order. However a substantial body of work has revealed that
for a Higgs mass larger than about $72$ Gev there is a smooth
crossover instead of a sharp transition\cite{csikor}. The current
LEP bound for the standard model Higgs mass $m_H \gtrsim 115$ Gev,
all but rules out the possibility of a strong first order phase
transition  and suggests a smooth crossover from the broken symmetry
into the symmetric phase in the standard model. Supersymmetric
extensions of the standard model may accommodate a strong first
order phase transition, but we will focus our study on the standard
model in this article.

\vspace{2mm}

{\bf The goal of this article:} We study  the properties of
neutrino collective excitations at high temperatures but below the
symmetry breaking scale in the standard model. The focus is on
obtaining the dispersion relations and damping rates of collective
excitations of neutrinos in the medium by including the full
momentum dependence in the contribution from the vector bosons.

If, as strongly suggested by the numerical analysis\cite{csikor},
the standard model has a smooth crossover between the broken and the
unbroken symmetry phases, or even a second order phase transition,
the expectation value of the neutral component of the scalar doublet
diminishes monotonically as the symmetry breaking temperature scale
is approached from below. In this case the mass of the vector bosons
along with the (chiral symmetry breaking) masses for the leptons and
quarks vanish continuously as the symmetry restoration temperature
is approached from below.

Even when at finite temperature the non-abelian vector bosons
acquire a Debye (electric) mass $m_D \sim g T$ and a magnetic mass
$m_m \sim g^2 T$\cite{pisarski,lebellac,manuel}, an important
consequence is that there is a large abundance of vector bosons in
the medium which  enhances processes that are neglected at lower
temperatures.

 As it will be discussed in detail
below some of these processes lead to a damping rate of collective
excitations at lowest order in the weak coupling.

The propagation properties of neutrinos in a medium are obtained
from an evaluation of the self-energy which takes into account the
full propagator of the vector bosons up to one loop order in the
high temperature limit. The main assumptions that are used in our
study are the following:

\begin{itemize}
\item{ We consider standard model neutrinos, namely left handed
massless neutrinos with flavor number conservation. The study of
neutrino mixing and oscillations will be presented
elsewhere\cite{proximo}. }

\item{We study the high temperature regime  below the symmetry
breaking electroweak scale, $T_{EW}$  in which the masses of the
vector bosons $M_{W,H}(T)<<T<T_{EW}$. The temperature dependence
of the vector boson mass arises from the temperature dependence of
the expectation value of the neutral scalar in the standard model
scalar doublet. Under the assumption of a smooth crossover, as
seems to be indicated by the lattice data, or of a second order
transition, this expectation value monotonically diminishes as $T
\rightarrow T_{EW}$ from below. Since leptons and quarks obtain
masses $m_f(T)$ via Yukawa couplings to the same scalar that gives
the mass to the vector bosons the ratios remain almost constant,
namely $m_f(T)/M_{W,H}(T)\sim m_f(0)/M_{W,H}(0)\ll 1$. We neglect
possible logarithmic variations of the weak and Yukawa couplings
with temperature via the standard renormalization group running.
Therefore at high temperature we will \emph{neglect} the fermion
masses in our analysis and will consider  neutrinos and leptons to
be massless. }

\item{We also consider a CP symmetric thermal bath, thereby neglecting
contributions which are proportional to the baryon asymmetry of the
Universe.  }

\end{itemize}

\vspace{2mm}

{\bf Brief summary of main results:} The main results obtained here
are the following:

\begin{itemize}
\item{We obtain the effective Dirac equations of neutrinos in the medium from a linear response
approach. Under the assumption that in the standard model there is
either a smooth crossover or a second order transition, we consider
the regime of temperatures near but below the electroweak scale
within which the mass of the vector bosons is much smaller than the
temperature. In this regime we implement the hard thermal loop (HTL)
approximation\cite{pisarski,lebellac,weldon} and obtain the
dispersion relations of neutrino collective excitations in the
medium. It is argued that perturbation theory and the (HTL)
approximation are reliable for $T\gg M_{W,Z}(T)\gg gT$. }

\item{In this regime, the properties of the neutrino collective modes are
qualitative and quantitatively different from those of the fermionic
excitations in QCD and QED\cite{klimov,weldon,pisarski,lebellac} and
depend on two parameters: a chirally symmetric mass scale
$m_\nu=gT/4$ and the dimensionless combination $\Delta=M^2_W(T)/2\,
m_\nu T$. Collective excitations are left handed negative helicity
quasiparticles and left handed positive helicity quasiholes and
their respective antiparticles. For $\Delta < \Delta_c =
1.275\cdots$ there are two gapped branches of quasiparticle
excitations and one gapless and two gapped branches of quasihole
excitations. The lower quasiparticle branch and \emph{all} the
quasihole branches terminate at end point values of the momentum $k$
determined by $\Delta$ while the dispersion relation of the upper
gapped quasiparticle branch merges with the free field dispersion
relation for $k\gg m_\nu$. For $\Delta_c < \Delta < \pi/2$ the
quasiparticle spectrum features a pitchfork bifurcation and for
$\Delta > \pi/2$ there are no quasihole branches available and only
a gapless quasiparticle branch remains. Its dispersion relation is
below the light cone for soft momenta $k \ll m_\nu$ and approaches
the free field limit for $k \gg m_\nu \,\Delta$ with a sharp
crossover between the non-perturbative and perturbative regimes at
$k \sim m_\nu$. }

\item{Quasiparticles and quasiholes acquire a width as a result of the
\emph{decay} of vector bosons into neutrinos and leptons. This is a
novel feature which distinguishes these excitations from those of
high temperature QCD and QED and has a simple kinetic
interpretation. We obtain the damping rate for quasiparticle and
quasiholes both at rest as well as for fast moving excitations with
$k \gg m_\nu\,\Delta$. These widths are of $\mathcal{O}(g^2)$ and it
is argued that they can be larger than the collisional width of
$\mathcal{O}(G^2_F)$\cite{notzold,tututi,thomson} at high
temperature for momenta $k \lesssim M_W(T)$. }

\item{The domain of validity of the one-loop (HTL) approximation as
well as the gauge invariance of the results are discussed in
detail.}

\end{itemize}

The article is organized as follows. In section (\ref{direqn}) we
obtain the effective Dirac equation for neutrinos in the medium
implementing a real-time approach in linear response. In section
(\ref{selfenergy}) we study the one-loop self energy in the (HTL)
approximation. In section (\ref{disp}) the   spectrum of collective
excitations and their damping rates are obtained. In this section we
discuss the regime of validity of the (HTL) approximation and give a
simple kinetic argument that shows that the decay of vector bosons
implies a  width for the collective excitations. Section
(\ref{disc}) is devoted to a discussion of the results and their
gauge invariance as well as a comparison between the collisional
damping rate of $\mathcal{O}(G^2_F)$ and the one-loop damping rate.
Our conclusions and further questions are summarized in section
(\ref{conc}). An appendix provides the technical details for the
self-energy.

\section{The effective Dirac equation in the medium}\label{direqn}

Anticipating a study of the real time dynamics of neutrino
propagation in a medium, including eventually mixing and
oscillations\cite{proximo}, we obtain the real-time effective Dirac
equations for neutrinos propagating in a thermal medium. While in
this article we focus on standard model neutrinos, the
generalization to include mixing, sterile, right handed or Majorana
neutrinos is straightforward.

The computation of the self-energy is carried out in the unitary
gauge. This gauge reveals the correct physical degrees of
freedom\cite{peskin} and previous calculations of the neutrino
self-energy in covariant gauges (one of which is the unitary gauge)
have proven the gauge independence of the dispersion
relations\cite{dolivoDR}.

The  part of the standard model Lagrangian density which is relevant
to our discussion is the following

\be\label{LSM} \mathcal{L}_{SM} =  \overline{\nu}_a \left(i
{\not\!{\partial}}\right) L  \nu_a +
\mathcal{L}^0_{W}+\mathcal{L}^0_Z +
\mathcal{L}_{CC}+\mathcal{L}_{NC}\ee

\noindent where $\mathcal{L}^0_{W,Z}$ are the free field Lagrangians
for the vector bosons in unitary gauge, namely

\bea \mathcal{L}^0_{W} & = & -\frac{1}{2}
\left(\partial_{\mu}W^+_{\nu}-\partial_{\nu}W^+_{\mu}
\right)\left(\partial^{\mu}W^{-\,\nu}-\partial^{\nu}W^{-\,\mu}
\right)+ M^2_{W}W^+_{\mu}W^{-\,\mu} \label{LW}\\
\mathcal{L}^0_{Z} & = & -\frac{1}{4}
\left(\partial_{\mu}Z_{\nu}-\partial_{\nu}Z_{\mu}
\right)\left(\partial^{\mu}Z^{\nu}-\partial^{\nu}Z^{\mu} \right)+
\frac{1}{2}M^2_{Z}Z_{\mu}Z^{\mu} \label{LZ}\eea

\noindent and the charged and neutral current interaction Lagrangian
densities are given by

\be \mathcal{L}_{CC} = \frac{g}{\sqrt{2}} \left[
\overline{\nu}_a\gamma^\mu L \, l_a W^+_{\mu} + \overline{l}_a
\gamma^\mu L \nu_a W^-_{\mu} \right]\label{LCC} \ee

\be \mathcal{L}_{NC} = \frac{g}{2 \cos \theta_w} \left[
\overline{\nu}_a\gamma^\mu L \nu_a Z_{\mu} + \overline{f}_a
\gamma^\mu (g^V_a-g^A_a\,\gamma^5) f_a Z_{\mu} \right]\label{LNC}\ee

In the above expressions $L=(1-\gamma^5)/2$ is the projection
operator on left handed states, $a$ is a flavor index and $g^{V,A}$
are the vector and axial vector couplings of quarks and leptons, $l$
stand for leptons and $f$ generically for the fermion species with
neutral current interactions.

The effective Dirac equation in  the medium in real time is derived
in linear response  by implementing the methods of non-equilibrium
quantum field theory described in ref.\cite{nosfermions}. In this
approach an external source induces an expectation value for the
neutrino field and this expectation value obeys the effective
equation of motion in the medium, including self-energy
contributions\cite{nosfermions}. This approach allows the study of
the real-time dynamics as an \emph{initial value problem} which will
be a distinct advantage to study of neutrino oscillations in the
medium.

To implement this approach we introduce an external Grassmann-valued
source that couples linearly to the neutrino field via the
Lagrangian density

\be \mathcal{L}_S = \overline{\nu}_a \eta_a + \overline{\eta}_a
\nu_a \label{Lsource} \ee

In presence of this source term, the total Lagrangian density is
given by $\mathcal{L}_{SM}+\mathcal{L}_S$.

The calculation of \emph{expectation values} (rather than in-out
S-matrix elements) requires the generating function of real time
correlation functions, which involves evolution along forward and
backward time branches\cite{ctp}. A path integral representation of
this generating functional is given by

\be\label{ZJ} \mathcal{Z}[j^+,j^-] = \int D\Phi^+ \, D\Phi^-
e^{i\int (\mathcal{L}[\Phi^+,j^+]-\mathcal{L}[\Phi^-,j^-])} \ee

\noindent where we have generically denoted by $\Phi^{\pm}$
\emph{all} the fields defined along the forward $(+)$ and backward
$(-)$ time branches to facilitate the notation. The sources
$j^{\pm}$ are linearly coupled to these fields in order to obtain
the real time correlation functions as functional derivatives of the
generating functional with respect to these sources. While the
sources $j^{\pm}$ are introduced to obtain the real time correlation
functions and are set to zero after the calculations, the external
(Grassman) source $\eta$ is the same along both branches $\pm$ and
induces an expectation value of the neutrino field which is the same
along both branches. See
references\cite{nosfermions,ctp,disip,tadpole} for further
discussion. The equation of motion for the expectation value of the
neutrino field induced by the external source is obtained by
shifting the field

\be \nu^{\pm}_a = \chi_a + \Psi^{\pm}_a ~~;~~ \chi_a = \langle
\nu^{\pm}_a \rangle ~~;~~ \langle \Psi^{\pm}_a \rangle =0
\label{shift}\ee

The equation of motion for the expectation value is obtained by
requesting that the $\langle \Psi^{\pm}_a \rangle =0$ be fulfilled
order by order in perturbation theory\cite{disip,tadpole}.
Implementing this method up to one-loop order we find the following
equation of motion\cite{nosfermions}

\be i\left(\not\!{\partial}\right) L \, \chi_a (\vx,t) + \int d^3 x'
\int dt' \Sigma^{ret}_{ab}(\vx-\vx',t-t')\chi_b(\vx',t') = -
\eta_a(\vx,t) \label{eqnofmot}\ee

\begin{figure}[h!]
\begin{center}
\includegraphics[height=3in,width=4in,keepaspectratio=true]{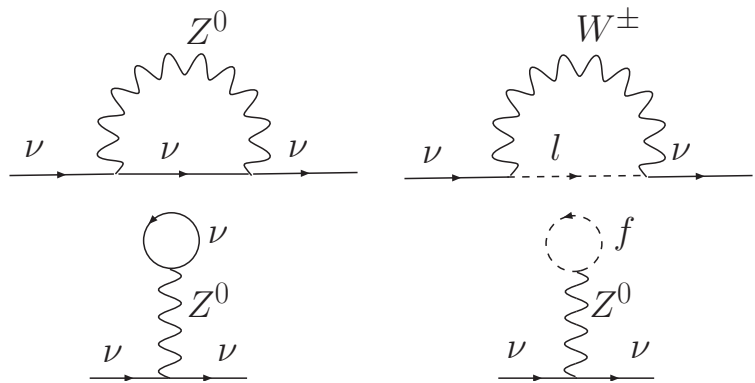}
\caption{One loop diagrams contributing to the neutrino self-energy.
The tadpole diagrams (neutral currents) with neutrino and lepton
loops are not considered in the case of a CP symmetric medium.}
\label{neusigma}
\end{center}
\end{figure}

Where $\Sigma^{ret}_{ab}(\vx-\vx',t-t')$ is the retarded
self-energy in real time whose one loop contributions are
displayed in fig. (\ref{neusigma}). In this study we will consider
only a CP symmetric medium, therefore we neglect the neutral
current one-loop tadpole diagrams with a closed neutrino or lepton
loop and only the one loop diagrams with the exchange of charged
or neutral vector bosons is considered, namely the two top
diagrams in  figure (\ref{neusigma}).

The one loop contributions from vector boson exchange to the
retarded self-energy are computed in the appendix. It is convenient
to introduce the space-time Fourier transform of the expectation
value and the source

\be \chi_a (\vx,t) = \frac{1}{\sqrt{V}}\sum_{\vk}\int d\omega \,
\chi_a (\vk,\omega) e^{i\vk\cdot\vx}\,e^{-i\omega\,t}~~;~~\eta_a
(\vx,t) = \frac{1}{\sqrt{V}}\sum_{\vk}\int d\omega \, \eta_a
(\vk,\omega) e^{i\vk\cdot\vx}\,e^{-i\omega\,t}\label{FTs}\ee

\noindent where we have kept the same name for the variables to
avoid introducing further notation.

Since we are considering the high temperature limit with $T$ much
larger than the lepton  masses (only the lightest leptons are
considered) we neglect the fermion masses in the computation of
the self energy. As it will become clear from the discussion
below, this is warranted because, in the high temperature limit
the loop momentum is of order $T$. In this case the self-energies
are proportional to the identity matrix in  flavor space. Using
the results of the appendix, the equation of motion in frequency
and momentum for each individual flavor becomes

\be \left[\gamma^0\,\omega- \vec{\gamma}\cdot \vk +
\Sigma_W(\vk,\omega)+ \Sigma_Z(\vk,\omega) \right]L\,\chi_a
(\vk,\omega)=-\eta_a(\vk,\omega)\,. \label{EqMotFT}\ee

The one loop contribution to the self-energies
$\Sigma_{W,Z}(\vk,\omega)$  are given by the first two diagrams in
fig.(\ref{neusigma}). The details of their calculation is presented
in the appendix and their final expressions are   given by equations
(\ref{retSEfin}) in terms of  the dispersive forms
(\ref{retSWdisp},\ref{retSZdisp}).

\section{The self energy: charged and neutral
currents}\label{selfenergy}

From the results of the appendix, in particular equations
(\ref{retSEfin}) and (\ref{imparts}) we obtain the following
explicit expression for the self-energies:

\be  \Sigma_{W,Z}(\vk,\omega) = \int \frac{dk_0}{\pi}
\frac{\textit{Im}\,{\Sigma}_{W,Z}(\vk,k_0)}{k_0-\omega-i\epsilon}
\label{retSEfin2} \ee

\bea \textit{Im}\,{\Sigma}_{W}(\vk,k_0) & = &  \frac{g^2\pi}{2} \int
\dbarq \frac{1}{4W_q\,p}\Bigg\{\left[
1-N_f(p)+N_b(W_q)\right]\left[\not\!{Q}(\vp,\vq)\,
\delta(k_0-p-W_q)+\not\!{Q}(-\vp,-\vq)\, \delta(k_0+p+W_q)\right]\nonumber \\
& + & \left[N_f(p)+N_b(W_q) \right]\left[\not\!{Q}(\vp,-\vq)\,
\delta(k_0-p+W_q)+\not\!{Q}(-\vp,\vq)\,
\delta(k_0+p-W_q)\right]\Bigg\}\label{ImsigW}\eea

\noindent where

\bea Q^{\mu}(\vp,\vq) & = & p^\mu+ 2q^\mu \Bigg(\frac{W_q\,p-
\vq\cdot\vp}{M^2_W(T)}\Bigg) \label{bigQ}\\
p^\mu & = & (\,|\vk-\vq |\,,  \vk-\vq\,) ~~;~~ q^\mu = (\,W_q, \vq\,) \label{pq}\\
W_q & = & \sqrt{q^2+M^2_W(T)}\,. \label{Wq}\\
\eea

The contribution from neutral currents is obtained from the above
expression by the replacement

\be \frac{g}{\sqrt{2}} \rightarrow \frac{g}{2\cos\theta_w}~~;~~
M_W(T)\rightarrow M_Z(T)=
\frac{M_W(T)}{\cos\theta_w}\label{repla}\ee

\noindent in what follows we will use the standard model value
$\sin^2\theta_w=0.23$ for numerical analysis, neglecting possible
finite temperature corrections.

The delta functions in the expression for (\ref{ImsigW}) (and
similarly for the neutral current contribution) have a clear
interpretation: those of the form $\delta(p+W_q \mp k_0)$
correspond to the process of \emph{decay} of a neutrino (positive
or negative energy) into an (anti) lepton and a vector boson and
the inverse process of recombination. These processes remain at
zero temperature but they imply a production threshold that can
\emph{only} be satisfied if the neutrino features a \emph{gap} in
its spectrum with a value larger than the mass of the vector
bosons. The delta functions of the form $\delta(W_q-p\mp k_0)$
correspond to the processes of vector boson decay into a
neutrino-(anti) lepton pair and its inverse (recombination)
process. These processes are only available in the medium and
their contribution vanishes in the zero temperature limit.

The form of the imaginary part of the self energies suggests that
the full self-energies can be written as follows

\be {\Sigma}_{W,Z}(\vk,\omega)= \gamma^0
\sigma^0_{W,Z}(k,\omega)-\vec\gamma\cdot
\widehat{\bf{k}}\,\sigma^1_{W,Z}(k,\omega)\label{selfes}\ee

The corresponding scalar functions
$\sigma^0_{W,Z}(k,\omega),\sigma^1_{W,Z}(k,\omega)$ are obtained
by projection. Their explicit expression is given by

\bea \textit{Im}\,{\sigma}^0_{W}(k_0,k) & = &  \frac{g^2\pi}{2} \int
\dbarq \frac{1}{4W_q\,p}\Bigg\{\left[
1-N_f(p)+N_b(W_q)\right]{Q}_{0}(\vp,\vq)\,\left[
\delta(k_0-p-W_q)+ \delta(k_0+p+W_q)\right]\nonumber \\
& + & \left[N_f(p)+N_b(W_q) \right]{Q}_{0}(\vp,-\vq)\,\left[
\delta(k_0-p+W_q)+
\delta(k_0+p-W_q)\right]\Bigg\}\label{Imlisig0W}\eea

\bea \textit{Im}\,{\sigma}^1_{W}(k_0,k) & = &  \frac{g^2\pi}{2} \int
\dbarq \frac{1}{4W_q\,p}\Bigg\{\left[
1-N_f(p)+N_b(W_q)\right]\,\uvk\cdot\vec{Q}(\vp,\vq) \,\left[
\,\delta(k_0-p-W_q)-\delta(k_0+p+W_q)\right]\nonumber \\
& + & \left[N_f(p)+N_b(W_q) \right]\,\uvk \cdot\vec{Q}(\vp,-\vq)
\,\left[\,\delta(k_0-p+W_q)-
\,\delta(k_0+p-W_q)\right]\Bigg\}\label{Imlisig1W}\eea

Using the dispersive representation (\ref{retSEfin2}) we find the
following expressions

\bea \label{sig0W} {\sigma}^0_{W}(\omega,k) & = & \frac{g^2}{2} \int
\dbarq \frac{1}{4W_q\,p}\Bigg\{\left[
1-N_f(p)+N_b(W_q)\right]{Q}_{0}(\vp,\vq)\left[\frac{1}{p+W_q-\omega-i\epsilon}
-\frac{1}{p+W_q+\omega+i\epsilon} \right]\nonumber \\
& + & \left[N_f(p)+N_b(W_q)
\right]{Q}_{0}(\vp,-\vq)\left[\frac{1}{p-W_q-\omega-i\epsilon}
-\frac{1}{p-W_q+\omega+i\epsilon} \right]\Bigg\}\eea

\bea \label{sig1W} {\sigma}^1_{W}(\omega,k) & = & \frac{g^2}{2} \int
\dbarq \frac{1}{4W_q\,p}\Bigg\{\left[
1-N_f(p)+N_b(W_q)\right]\,\uvk\cdot\vec{Q}(\vp,\vq)
\,\left[\frac{1}{p+W_q-\omega-i\epsilon}
+\frac{1}{p+W_q+\omega+i\epsilon} \right]\nonumber \\
& + & \left[N_f(p)+N_b(W_q)
\right]\,\uvk\cdot\vec{Q}(\vp,-\vq)\,\left[\frac{1}{p-W_q-\omega-i\epsilon}
+\frac{1}{p-W_q+\omega+i\epsilon} \right]\Bigg\}\eea

The expressions for the neutral current contributions can be
obtained by the simple replacement (\ref{repla}) and adding both
contributions defines the following scalar functions

\be {\sigma}^{0}(\omega,k) =
{\sigma}^{0}_W(\omega,k)+{\sigma}^{0}_Z(\omega,k)~;~~{\sigma}^{1}(\omega,k)
= {\sigma}^{1}_W(\omega,k)+{\sigma}^{1}_Z(\omega,k)\label{sigmas}\ee

An important property of the imaginary parts that will be useful in
the analysis of the width of the collective modes is the following

\bea \textrm{Im}\,\sigma^0(-\omega,k) & = &
\textrm{Im}\,\sigma^0(\omega,k)\label{evenimsig0}\\
\textrm{Im}\,\sigma^1(-\omega,k) & = &-
\textrm{Im}\,\sigma^1(\omega,k).\label{oddimsig1}\eea

\subsection{Hard thermal loops }

In the high temperature limit $T\gg M_{W,Z}(T)$ the integrand  in
the expressions above are dominated by loop momentum $q \sim T$.
This is simply gleaned from the high powers of momentum that
multiply the Fermi-Dirac and Bose-Einstein distribution functions.
In this limit, the hard thermal loop (HTL)
approximation\cite{pisarski,lebellac,weldon} is warranted. For
frequency and momentum of the neutrino excitations $k,\omega \ll T$
the (HTL) ``counting'' $q\sim T \gg k,\omega$ leads to the following
approximations

\be
 p \sim q-\vk \cdot \uvq ~~;~~ W_q \sim q + \frac{M^2}{2q}\,.
 \label{htlappx} \ee

\noindent where $M$ stands generically for $M_{W,Z}(T)$.

 Ignoring the vacuum contribution, a lengthy but straightforward calculation using these
approximations leads to the following results for the real parts

\bea \textrm{Re}\,  {\sigma}^0_{W}(k,\omega)  & = & \frac{m^2_\nu
\,\omega}{12\,M^2_W(T)} - \frac{m^2_\nu}{2k}\, I(\omega,k,\Delta)
\label{Resig0W} \\
\textrm{Re}\,  {\sigma}^1_{W}(k,\omega) & = & -\frac{m^2_\nu
\,k}{18\,M^2_W(T)}  + \frac{m^2_\nu}{2k}\,J(\omega,k,\Delta)
\label{Resig1W} \eea

\noindent where

\be m_\nu = \frac{g\,T}{4} \label{numass} \ee

The dimensionless functions $I(\omega,k)$ and $J(\omega,k)$ are more
compactly expressed by introducing the dimensionless variables

\bea \overline{\omega} &  =  & \frac{\omega}{m_\nu} ~~;~~
\overline{k}   =   \frac{k}{m_\nu}\label{dimvars} \\
\Delta  & = & \frac{M^2_W(T)}{2 \, m_\nu \, T} = \frac{g}{8}
\left(\frac{M_W(T)}{m_\nu} \right)^2\label{Delta}\eea

\noindent and the functions

\bea  LP(\overline{\omega},\overline{k},\Delta;z)  & = & \frac{1}{2}
\ln\Bigg|\frac{\bw+\bk+\frac{\Delta}{z}}{\bw-\bk+\frac{\Delta}{z}}
\Bigg| \label{LpW} \\
LM(\overline{\omega},\overline{k},\Delta;z)  & = &\frac{1}{2}
\ln\Bigg|\frac{\bw+\bk-\frac{\Delta}{z}}{\bw-\bk-\frac{\Delta}{z}}
\Bigg| \label{LmW}\eea

In terms of which

\be  \label{IW} I(\omega,k,\Delta)  =  \frac{4}{\pi^2} \int^\infty_0
dz ~\frac{2z\,e^{-z}}{1-e^{-2z}}\left[
LP(\overline{\omega},\overline{k},\Delta;z)+
LM(\overline{\omega},\overline{k},\Delta;z) \right] \ee

\be  \label{JW} J(\omega,k,\Delta)  =  \frac{4}{\pi^2} \int^\infty_0
dz ~\frac{2z\,e^{-z}}{1-e^{-2z}}\left[2-
\,\frac{\bw+\frac{\Delta}{z}}{\bk}\,
LP(\overline{\omega},\overline{k},\Delta;z)-\frac{\bw-\frac{\Delta}{z}}{\bk}
\,LM(\overline{\omega},\overline{k},\Delta;z) \right] \ee

Similar expressions are obtained for the neutral current
contributions by the replacement (\ref{repla}). Adding together the
charged and neutral current contributions leads to

\bea I_T(\omega,k,\Delta)  & = &
I(\omega,k,\Delta)+\frac{1}{2\cos^2\theta_w}\,I\left(\omega,k,\frac{\Delta}{\cos^2\theta_w}\right)\label{Itot}\\
J_T(\omega,k,\Delta)  & = &
J(\omega,k,\Delta)+\frac{1}{2\cos^2\theta_w}\,J\left(\omega,k,\frac{\Delta}{\cos^2\theta_w}\right)\label{Jtot}\eea

 We note that for $M_W=0$

\be I(\omega,k,0)=\ln\Bigg|\frac{\omega+k}{\omega-k}
\Bigg|\label{Ihtl}\ee

\be J(\omega,k,0) =2 -
\frac{\omega}{k}\ln\Bigg|\frac{\omega+k}{\omega-k}
\Bigg|\label{Jhtl}\ee

\noindent which are the standard results of the (HTL) approximation
for vector-like theories without spontaneous symmetry breaking
(QCD-QED)\cite{weldon,pisarski,lebellac}. The following properties
of these functions will be important in the discussion of the
collective modes below,

\bea I(-\omega,k,\Delta) & = & -I(\omega,k,\Delta) \label{woddI}\\
J(-\omega,k,\Delta) & = & J(\omega,k,\Delta) \label{wevenJ}\\
I(\omega,-k,\Delta) & = & - I(\omega,k,\Delta) \label{koddI}\\
J(\omega,-k,\Delta) & = & J(\omega,k,\Delta) \label{koddJ} \\
J(\omega,0,\Delta) & = & 0 \label{Jk00} \eea

\section{Dispersion relations and widths of  collective
excitations}\label{disp}

The form of the self-energies (\ref{selfes})  allow to write the
\emph{homogeneous} Dirac equation (setting to zero the external
Grassmann sources $\eta$ ) in the form

\be
\left[\Lambda_+(\uvk)D_-(\omega,k)+\Lambda_-(\uvk)D_+(\omega,k)\right]\,L
\chi_a(\vec{k},\omega)= 0 \label{Direqn}\ee

\noindent where

\be \Lambda_{\pm}(\uvk) = \frac{1}{2}(\gamma^0 \mp \,
\vec{\gamma}\cdot\uvk) = \frac{\gamma^0}{2}(1\mp
h(\uvk)\,\gamma^5) \label{pros}\ee

\noindent here $h(\uvk)$ is the helicity operator, and

\bea D_+(\omega,k)  & = &  \omega-k +
\sigma^0(k,\omega)-\sigma^1(k,\omega) \label{Dplus}\\
D_-(\omega,k)  & = &  \omega+k +
\sigma^0(k,\omega)+\sigma^1(k,\omega)\label{Dminus} \eea

Therefore, the propagator for left handed fields is given by

\be \label{propaga} iS(\omega,k) =
\frac{\Lambda_+(\uvk)}{D_+(\omega,k)}+\frac{\Lambda_-(\uvk)}{D_-(\omega,k)}\ee

The neutrino spectral function is determined from the propagator
above, and is given by

\be \label{rho} \rho(\omega,k)   =   -\frac{1}{\pi}
\textrm{Im}[iS(\omega+i\epsilon,k)] =
\Lambda_+(\uvk)\rho_+(\omega,k)+  \Lambda_-(\uvk)\rho_-(\omega,k)
\ee

\noindent where

\be   \rho_\pm(\omega,k)  =
\frac{1}{\pi}\frac{\textrm{Im}\,D_\pm(\omega+i\epsilon,k)}{\Big[\textrm{Re}\,D_\pm(\omega+i\epsilon,k)
\Big]^2+\Big[\textrm{Im}\,D_\pm(\omega+i\epsilon,k) \Big]^2}
\label{rhopm}\ee

The poles of the propagator along the real axis in the complex
$\omega$ plane correspond to physical stable excitations, while
complex poles very near the real axis in a second (or higher)
Riemann sheet in this plane, describes quasiparticles or
resonances. Propagating quasiparticles are characterized by narrow
width resonances, namely their widths must be much smaller than
the real part of the (complex) pole, so the decay rate of the
quasiparticle is much smaller than its oscillation frequency.

The position of the (quasi) particle poles  are $\omega_{\pm}(k)$
which are  determined by the conditions

\bea  \textrm{Re}\,D_+(\omega_+(k),k)   & = &  0
\label{ppoles}\\
\textrm{Re}\,D_-(\omega_-(k),k)   & = &  0 \label{mpoles} \eea

\emph{If} these functions vanish \emph{linearly} near their zeroes
then near the resonances the spectral density can be approximated by
a Breit-Wigner form, namely

\be \rho_\pm(\omega,k) \sim \frac{Z_\pm(k)}{\pi}
\frac{\Gamma_\pm(k)}{\Big(\omega-\omega_\pm(k)\Big)^2+
\Gamma^2_\pm(k)}\label{BW}\ee

\noindent where the residues $Z_\pm(k)$ and widths (damping rates)
$\Gamma_\pm(k)$ are given by

 \be Z^{-1}_\pm(k) = \Bigg|\frac{\partial D_\pm(\omega,k)}{\partial
 \omega}\Bigg|_{\omega=\omega_\pm(k)} \label{Zpm}\ee

 \be \Gamma_\pm(k) = Z_\pm(k) \, \textrm{Im}\,D_\pm(\omega_\pm(k),k)
 \label{widths}\ee

The spinor wave functions of the resonances (quasiparticles) are
obtained from the homogeneous effective Dirac equation in the
medium (\ref{Direqn}) by considering only the real part of
$D_\pm(\omega,k)$, since the vanishing of the real parts defines
the quasiparticles. Because  neutrinos are left handed  fields the
eigenspinors corresponding to the solutions of
$D_+(\omega_+(k),k)=0$ must obey

 \be (1+ h(\uvk))L \chi(\omega_+(k),k) =0 \label{solplus}\ee

 Namely the spinor eigenstate is a left handed, negative helicity
 state, just as the usual neutrino field. We will refer to the \emph{positive energy} solutions of $D_+(\omega_+(k),k)=0$
  as \emph{quasiparticles}.
 Similarly, the spinors solutions corresponding to the roots of $D_-(\omega_-(k),k)=0$ must
 obey

 \be (1- h(\uvk))L \chi(\omega_-(k),k) =0 \label{solmin}\ee

Corresponding to a  left handed neutrino state with \emph{positive}
helicity. Following ref.\cite{weldon}, we will refer to the
\emph{positive energy} solutions of $D_-(\omega_-(k),k)=0$ as
\emph{quasiholes}. These states feature the \emph{opposite}
chirality-helicity assignment of the vacuum neutrino states.

As it will be discussed in detail in section (\ref{domain}), the
regime of validity of the perturbative expansion is restricted to
$M_{W,Z}(T) \gg g\,T$ in which case $m_\nu/M_{W,Z}(T) \ll 1$.
Therefore  the first terms in equations
(\ref{Resig0W},\ref{Resig1W})  provide a perturbative correction
to the coefficients of $\omega$ and $k$ in $D_{\pm}(\omega,k)$.
Since the focus is to explore the non-perturbative aspects of
``soft'' collective excitations we will ignore these terms in what
follows. The final expressions for $ D_\pm(\omega,k)$ are given by

\be \,D_+(\omega,k) = \omega-k - \frac{m^2_{\nu}}{2k}\left[
I_T(\omega,k)+J_T(\omega,k)\right]+ i\left[ \textit{Im}\,{\sigma}^0
(\omega,k)-\textit{Im}\,{\sigma}^1(\omega,k)\right]
\label{Dplusri}\ee

\be \,D_-(\omega,k) = \omega+k - \frac{m^2_{\nu}}{2k}\left[
I_T(\omega,k)-J_T(\omega,k)\right]+i\left[
\textit{Im}\,{\sigma}^0(\omega,k)+\textit{Im}\,{\sigma}^1(\omega,k)\right]
\label{Dminri}\ee

The real and imaginary parts of $D_{\pm}(\omega,k)$ feature the
following symmetries

\bea \textit{Re}\, D_+(-\omega,k) = - \textit{Re}\,
D_-(\omega,k)\label{symreal}\\
\textit{Im}\, D_+(-\omega,k) =  \textit{Im}\,
D_-(\omega,k)\label{symimag}\eea

The dispersion relations of quasiparticles and quasiholes are
obtained from  roots of the real part of the self-energy, namely

\be \textit{Re}\,D_{\pm}(\omega_{\pm}(k),k)  =  0 \label{roots}\ee

 The symmetry relation
(\ref{symreal}) indicates that we only need to find the
\emph{positive} roots in eqn. (\ref{roots}), therefore the roots of
$\textit{Re}\,D_{\pm}(\omega,k)=0$ will be the pairs
$(\omega_+(k),-\omega_-(k));(\omega_-(k);-\omega_+(k))$
respectively, with  $\omega_{\pm}(k)\geq 0$.

Antiquasiparticles and antiquasiholes have negative energy and
quantum numbers opposite to those of quasiparticles and
quasiholes. The interpretation of the excitation spectrum  is as
follows\cite{weldon}: the positive energy roots of $D_+(\omega,k)$
correspond to negative helicity left handed \emph{quasiparticle}
states of energy $\omega_+(k)$ and spectral weight $Z_+(k)$ while
the negative energy roots are left handed \emph{anti-quasihole}
states of negative helicity of energy $-\omega_-(k)$ and residue
$Z_-(k)$. The positive energy roots of $D_-(\omega,k)$ correspond
to left handed positive helicity \emph{quasiholes} of energy
$\omega_-(k)$ and spectral weigth $Z_-(k)$ and the negative energy
roots correspond to left handed positive helicity
\emph{anti-quasiparticles} of energy $-\omega_+(k)$ and residue
$Z_+(k)$. The relation between the imaginary parts, eqn.
(\ref{symimag}) states that quasiparticles and antiquasiparticles
have the same width and so do quasiholes and antiquasiholes, this
is obviously a consequence of the underlying  CPT symmetry.

The table below summarizes the properties of the collective
excitations\cite{weldon}

\begin{center}
\begin{tabular}{|c|c|c|c|c|}
\hline
 Identity & E & h & Z & $\Gamma(k)$\\
\hline
 \textrm{quasiparticle}~ & ~$\omega_+(k)$ ~ & ~-1 ~& $ ~ Z_+(k)$ & $~\Gamma_+(k)$\\
\hline
 \textrm{antiquasiparticle}~ & ~ $-\omega_+(k)$ ~ & ~ 1 ~& ~$Z_+(k)$ & $~\Gamma_+(k)$ \\
\hline
 \textrm{quasihole}~ & ~ $\omega_-(k)$~ & ~ 1 & ~ $Z_-(k)$ ~& $~\Gamma_-(k)$ \\
\hline
\textrm{antiquasihole}~  &  ~ $-\omega_-(k)$~ &  ~ -1 & ~ $Z_-(k)$~& $\Gamma_-(k)$ \\
\hline
\end{tabular}\label{assign}
\end{center}

\subsection{Gaps in the  spectrum of collective modes}\label{gaps}
The position of the gaps in the spectrum of collective excitations
are determined by the roots of the equation

\be \textrm{Re}\,D_\pm(\omega,k=0,\Delta)=0 \label{gapeqn} \ee

\noindent where we have used the property $J(\omega,k=0)=0$ (see
eqn. (\ref{Jk00})) which implies that
$D_+(\omega,k=0,\Delta)=D_-(\omega,k=0,\Delta)$. General features of
the gap equation (\ref{gapeqn}) can be understood from the
expression

\be \label{Dk0} D_+(\omega,0,\Delta)=
m_\nu\,\left\{\overline{\omega} +
\frac{1}{2}\left[H(\bw,\Delta)+\frac{1}{2\cos^2\theta_w}H\bigg(\bw,\frac{\Delta}{\cos^2\theta_w}\bigg)\right]\right\}\ee

\noindent where

\be H(\bw,\Delta) = \frac{8\,\bw}{\pi^2} \int^\infty_0 dz
~\frac{2z\,e^{-z}}{1-e^{-2z}}\mathcal{P}\left(
\frac{1}{\frac{\Delta^2}{z^2}-\bw^2}\right) . \label{I0}\ee

\noindent and $\mathcal{P}$ stands for the principal part.

For $\Delta=0$, $H(\omega,0) = -2 \,\mathcal{P}(1/\omega)$ which
leads to the solution of the gap equation

\be \label{gapD0}\omega(k=0) = \pm\,m_\nu
\,\Big(1+\frac{1}{2\cos^2\theta_w}\Big)^{\frac{1}{2}}. \ee

We note that  because $\mathcal{P}(1/\omega)$ is an \emph{odd}
function of $\omega$ vanishing at $\omega=0$ (by definition of the
principal part) there are \emph{three} roots of the eqn. $
D_+(\omega,0,0)=0$: $\omega=0$ and  the two roots corresponding to
the value of the gap (\ref{gapD0}). The gapless excitations
associated with the root $\omega=0$ will be explored in subsection
(\ref{Dmin}) below for general values of $\Delta$ including
$\Delta=0$.

For $\Delta \neq 0$ the function $D_+(\omega=0,k=0,\Delta)=0$
because it is an  odd function of $\omega$. Furthermore, for $
\Delta \ll 1$ and $\omega \approx 0$,   it follows that

\be H(\omega,\Delta) \sim \frac{\omega}{\Delta^2}\ee

\noindent therefore, for $\Delta \neq 0$, $D_+(\omega,k=0,\Delta)$
vanishes at $\omega=0$,  is positive for $\omega > 0$ and rises
sharply from the origin with a slope $\propto 1/\Delta^2$. For small
$\Delta$ the function $H(\omega,\Delta)$ eventually becomes negative
for large enough $\omega$. This is  because the integrand is sharply
peaked at $z \sim 2$, therefore, for sufficiently small $\Delta$ and
sufficiently large $\omega$ the region in which the denominator is
negative dominates. Since for large $\omega$, $D_+(\omega,\Delta)
\approx \omega$ there must be yet another sign change and the
function $D_+(\omega,\Delta)$ must have at least \emph{three
zeroes}, namely three roots. One of the roots corresponds to
$\omega=0$ which determines a \emph{gapless} branch of collective
excitations, however for small $\Delta$ one of the roots must be at
$\overline{\omega} \sim \Delta$ while another root should be at
$\overline{\omega} \sim
(1+\frac{1}{2\cos^2\theta_w})^{\frac{1}{2}}$. Therefore for $\Delta
\ll 1$ there are \emph{three} branches, one emerging from the origin
is gapless and two other branches that feature a gap in their
dispersion relations.

\begin{figure}[h!]
\begin{center}
\includegraphics[height=3in,width=4in,keepaspectratio=true]{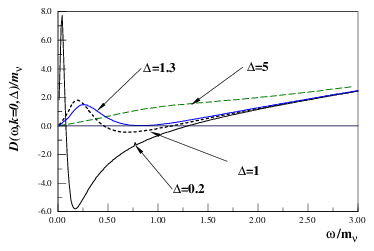}
\caption{Roots of $D(\omega,k=0,\Delta)$ for $\Delta=0.2,1,1.3,5$
respectively. The intersection of these curves with the horizontal
axis determines the values of the gap in the spectra of collective
modes. } \label{fig:gap}
\end{center}
\end{figure}

However for larger $\Delta$ the region in which the denominator
becomes negative necessarily corresponds to $z>>2$ and the integrand
is strongly suppressed in which case the function $H(\omega,\Delta)$
will be always positive and no root will be available. This behavior
is clearly displayed in fig. (\ref{fig:gap}) where we used the
standard model value $\sin^2\theta_w=0.23$.

This figure shows that there is a \emph{critical value} $\Delta_c$
such that  for $\Delta \lesssim \Delta_c$ there are \emph{two}
non-vanishing roots which coalesce at $\Delta=\Delta_c$  while for
$\Delta
> \Delta_c$ the only root  available is $\omega=0$. We find numerically that the
critical value is given by

\be\label{Deltac} \Delta_c \sim 1.275\cdots \ee

In order to resolve whether the gapless branch describes
quasiparticles or quasiholes, we must find the $k$ dependence of
the gapless branch.

A straightforward but lengthy calculation shows that to linear order
in $k$

\be D_\pm(\omega,k,\Delta) = D_+(\omega,0,\Delta) \mp
 k \,\Big[1+S(\bw,\Delta)+\frac{1}{2\cos^2\theta_w}S\big(\bw,\frac{\Delta}{\cos^2\theta_w}\big)\Big]\label{smaK}\ee

\noindent with $D_+(\bw,0,\Delta)$ given by eqn. (\ref{Dk0}) and

\be S(\bw,\Delta)= -\frac{4}{3\pi^2} \int^\infty_0 dz
~\frac{2z\,e^{-z}}{1-e^{-2z}}\mathcal{P}\Bigg[
\frac{\frac{\Delta^2}{z^2}+\bw^2}{\Big(\frac{\Delta^2}{z^2}-\bw^2
\Big)^2}\Bigg] \label{S}\ee

Therefore the dispersion relations for small $k$ near the gaps with
value $\omega_\pm(0)$ are given by

\be \omega_\pm(k)\sim \omega_\pm(0)+ c_\pm\, k + \mathcal{O}(k^2)
\label{displin} \ee

\noindent with the group velocities $c_\pm$   given by

\be c_\pm = \pm
\frac{1}{D'(\omega_\pm(0),0,\Delta)}\left[1+S(\bw_\pm(0),\Delta)+\frac{1}{2\cos^2\theta_w}S\big(\bw_\pm(0),\frac{\Delta}{\cos^2\theta_w}\big)
\right]\label{vg} \ee

Therefore for sufficiently small $k$ \emph{all} branches of
collective modes behave linearly with $k$. The study of the group
velocities for each branch necessarily has to be done numerically,
however of particular interest is the group velocity for
long-wavelength excitations of the \emph{gapless} branch. A
straightforward calculation for $\omega_\pm(0)=0$ leads to the
following result

\be c_{\pm} = \pm \Bigg[\frac{\Delta^2 -\frac{\pi^2}{4}}{\Delta^2 +
\frac{3\pi^2}{4}}\Bigg]\label{gaplessvg}\ee

\noindent where $\pm$ refer to the roots of $D_\pm(\omega,k,\Delta)$
respectively. As discussed above, quasiparticle and quasiholes are
\emph{positive energy} roots of $D_\pm(\omega,k,\Delta)$
respectively. Hence, the expression for the group velocity for long
wavelength excitations on the gapless branch  (\ref{gaplessvg})
indicates that for $\Delta < \pi/2$ positive energy excitations on
the gapless branch are \emph{quasiholes}, while for $\Delta > \pi/2$
gapless collective modes are \emph{quasiparticles}. Therefore the
gapless branch changes identity at $\Delta = \pi/2$.

In summary, the spectrum of collective excitations always features a
\emph{gapless} branch for any value of $\Delta$. For $\Delta <
\pi/2$ these gapless collective modes are quasiholes, and for
$\Delta > \pi/2$ they are quasiparticles.

  For generic $\Delta
<\Delta_c \sim 1.275\cdots$ the quasiparticle spectrum will feature
at least two gapped branches, while the quasihole spectrum will
feature two gapped and one gapless branch. For $\Delta \ll 1$ the
lowest gapped branch begins at $|\omega (0)| \sim \Delta$ while the
highest one begins at $|\omega(0)| \sim m_\nu
(1+\frac{1}{2\cos^2\theta_w})^\frac{1}{2}$. For $\Delta > \pi/2$ the
quasiparticle spectrum only features a gapless branch and no
branches of collective excitations remain for quasiholes.

The dispersion relations are linear for small momentum $k$ in each
branch. However, the gapless branch corresponds to roots of
$D_-(\omega,k,\Delta)$ for $\Delta <\pi/2$ while it describes the
\emph{only} available, gapless branch of $D_+(\omega,k,\Delta)$ for
$\Delta > \pi/2$. A detailed numerical analysis of the quasiparticle
and quasihole collective modes is presented below.

\subsection{Quasiparticle spectrum}
The dispersion relation  of quasiparticles corresponds to the
positive roots of $\textit{Re}\, D_+(\omega_+(k),k,\Delta)$. The
study of the dispersion relation in the full range of $k$ and
$\Delta$ must necessarily be carried out numerically. However,
before doing so, we can gain insight into the nature of the
dispersion relations by considering the simplified case of
$\Delta=0$. While this case is outside the domain of validity of
the perturbative expansion because of the restriction that
$M_{W,Z}(T) \gg m_\nu$ (see section (\ref{domain})), the study of
this case highlights several relevant aspects of the dispersion
relation and also serves as comparison to the more familiar
results in the literature in the case of unbroken gauge theories
in the (HTL) approximation\cite{weldon,pisarski,lebellac}.

In the case $\Delta=0$ the function $D_+(\omega,k,0)$ is familiar
from the (HTL) approximation for fermionic excitations in unbroken
vector-like gauge theories\cite{weldon,pisarski,lebellac}

\be D_+(\omega,k,0) =
m_\nu\Bigg\{\bw-\bk+\frac{C}{2\bk}\Big[\Big(1-\frac{\bw}{\bk}\Big)\ln\Big|
\frac{\bw+\bk}{\bw-\bk} \Big|+2\Big]\Bigg\}~~;~~C= 1+
\frac{1}{2\cos^2\theta_w} \label{DpD0}\ee

This function is negative for $\bw \sim 0$ but becomes positive for
$\bw \gg k$ and  features only one positive root for any value of
$k$. The gap is given by

\be \omega_g =
m_\nu\,\big(1+\frac{1}{2\cos^2\theta_w}\big)^\frac{1}{2}\label{gapHTL}\ee

The group velocity for long-wavelength excitations around this gap
is obtained from the general expression (\ref{vg}) by setting
$\Delta=0$ and the value of the gap (\ref{gapHTL}). A
straightforward analysis yields $c_+ = 1/3$.

 This situation is very  different for $\Delta
\neq 0$. The  discussion on the values of the gaps in the spectrum
of collective excitations in the previous subsection (\ref{gaps}) as
well as the dispersion relation of long wavelength excitations in
the gapless branch indicates that there are three distinct regions
of the parameter $\Delta$ with qualitatively different behavior.

\subsubsection{$\Delta < \Delta_c\sim 1.275\cdots$.}
In this region the quasiparticle spectrum features two gapped
branches and no gapless branch. We denote the dispersion relation
for the lowest branch by $\omega^+_<(k)$ and and that for the
higher branch $\omega^+_>(k)$ respectively.

For $\Delta \ll 1$ the gap for the lowest branch is
$\omega^+_<(0)\sim \Delta~m_\nu$ whereas the gap for the highest
branch  is $\omega^+_>(0) \sim m_\nu
\big(1+\frac{1}{2\cos^2\theta_w}\big)^{\frac{1}{2}}$ (see fig.
(\ref{fig:gap})). For increasing values of $k$ the roots
corresponding to the largest gap move to larger values of
$\omega$, while the roots corresponding to the lowest gap move to
smaller values of $\omega$. The spectrum for the lowest branch
terminates at an end point $k_e$ which  numerically is found to be
$k_e \lesssim \Delta/2$. For $k> k_e$ there are no further roots
corresponding to the lowest branch and only the highest branch is
available. For $k\gg m_\nu$ the dispersion relation along this
remaining branch is found to be

\be \omega_+(k) \sim k +
\frac{m^2_\nu}{k}\Big(1+\frac{1}{2\cos^2\theta_w} \Big) +\cdots
\label{largek}\ee

\begin{figure}[h!]
\begin{center}
\includegraphics[height=3in,width=4in,keepaspectratio=true]{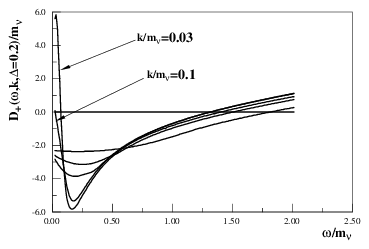}
\caption{Roots of $D_+(\omega,k,\Delta=0.2)/m_\nu$ for
$\frac{k}{m_\nu}=0.03,0.1,0.3,0.5,1$ respectively.}
\label{fig:rootsDpo2}
\end{center}
\end{figure}

The positive energy roots for $\Delta \ll 1$ are displayed in fig.
(\ref{fig:rootsDpo2}).

\begin{figure}[h!]
\begin{center}
\includegraphics[height=3in,width=4in,keepaspectratio=true]{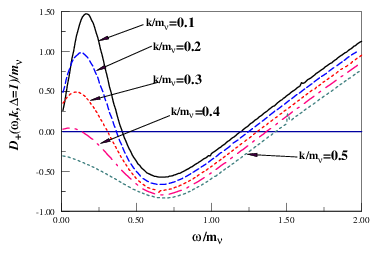}
\caption{Roots of $D_+(\omega,k,\Delta=1)/m_\nu$ for
$\frac{k}{m_\nu}=0.2,0.3,0.4,0.5$ respectively.}
\label{fig:rootsDplus}
\end{center}
\end{figure}

 For larger values of $\Delta < \Delta_c$ the behavior is
 qualitatively similar but with some quantitative differences.
The positive energy roots of $D_+(\omega,k,\Delta=1)$ for several
values of  $k$ are displayed in fig. (\ref{fig:rootsDplus}).

This figure clearly reveals that the branch with the largest gap
remains for any value of $k$, whereas that corresponding to the
smallest gap terminates at an end point value of momentum which
numerically is found to be roughly $k_e \lesssim \Delta/2$. For
$k>k_e$ there are no \emph{real} roots for the lowest branch.

The weight of the (weakly damped) collective modes to the spectral
function is determined by the residue $Z_+(k,\Delta)$ given by eqn.
(\ref{Zpm}) evaluated at the position of the root,
$\omega_+(k,\Delta)$, which determines the dispersion relation of
collective excitations on a particular branch.

 The analysis in the previous subsection shows that the function
 $D_+(\omega,k,\Delta)$ rises sharply for small $\omega$ with a slope
  $\sim 1/\Delta^2$ therefore for $\Delta \ll 1$ the weight of
 these collective modes with dispersion relation corresponding to
 the root $\omega^+_<(k,\Delta)$  is

 \be Z_+(k,\Delta) \sim \Delta^2 \ll 1 \label{resles}\ee

 Hence for $\Delta \ll 1$ the branch of collective modes with dispersion relation
 $\omega^+_<(k)$ has \emph{negligible} spectral weight.

We find that for $k\gg m_\nu$, the asymptotic behavior of the
collective modes in the upper branch is similar to the (HTL) limit
of unbroken vector-like theories\cite{weldon,pisarski,lebellac}.
For fast moving quasiparticles corresponding to the upper branch
in the limit  $k\gg m_\nu$ the dispersion relation is given by
eqn.  (\ref{largek}) and the spectral weight is found to be

\be Z_+(\omega_+(k),\Delta) \sim 1 -
\frac{m^2_\nu}{2k^2}\Big(1+\frac{1}{2\cos^2\theta_w} \Big)\ln\Big(
\frac{2k^2}{m^2_\nu}\Big)+\cdots\label{Zfast}\ee

\begin{figure}[h!]
\begin{center}
\includegraphics[height=2in,width=3in,keepaspectratio=true]{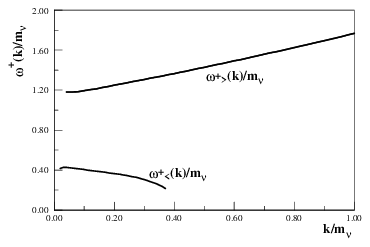}
\includegraphics[height=2in,width=3in,keepaspectratio=true]{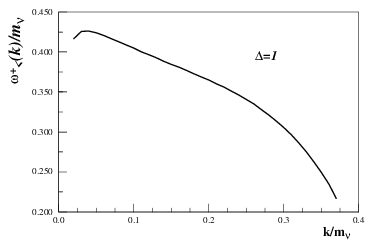}
\caption{Left: Dispersion relations for the higher ($\bw^+_>(k)$)
and lower ($\bw^+_<(k)$) branches of
$D_+(\omega,k,\Delta=1)/m_\nu$. Right: Lower branch in detail. The
lower branch terminates at a value $k_e(\Delta)$\,; $k_e(1)\sim
0.4$. } \label{fig:rutsDp}
\end{center}
\end{figure}

The dispersion relations for both branches together and for the
lowest branch separately are displayed in fig. (\ref{fig:rutsDp})
for $\Delta=1$.

\subsubsection{$\Delta_c<\Delta<\pi/2$: a pitchfork bifurcation of the spectrum. }\label{bifurca}

\begin{figure}[h!]
\begin{center}
\includegraphics[height=3in,width=4in,keepaspectratio=true]{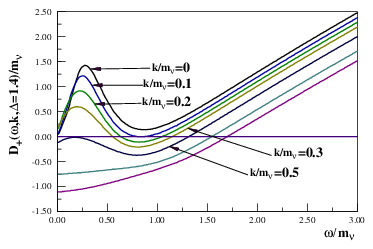}
\caption{Pitchfork bifurcation of the quasiparticle spectrum for
$\Delta > \Delta_c$. The function $D_+(\omega,k,\Delta=1.4)/m_\nu$
vs. $\omega/m_\nu$ for $k/m_\nu =0,0.1,0.2,0.3,0.5 $.  }
\label{fig:bifurca}
\end{center}
\end{figure}

As discussed above, for $\Delta > \Delta_c=1.275\cdots$ the only
solution of the gap equation is $\omega=0$, however for $\Delta <
\pi/2$ there is no gapless positive energy solution with linear
dispersion relation for small $k$ quasiparticles. Therefore the
following question arises, what is the nature of the quasiparticle
spectrum for $\Delta_c < \Delta < \pi/2$?.  A numerical study
reveals a remarkable answer:  for a given value of $\Delta$ in
this region there is \emph{critical} value of $k$, denoted by
$k_c(\Delta)$ for which  \emph{two roots} of the equation
$D_+(\omega,k,\Delta)=0$ emerge with a pitchfork bifurcation in
the spectrum. This result is depicted in fig. (\ref{fig:bifurca})
for $\Delta=1.4$.

Note in the figure (\ref{fig:bifurca}) that there are no roots for
$D_+(\omega,0,\Delta=1.4)$, however two roots emerge continuously
for $k > k_c \sim 0.1 ~m_\nu$.

For $\Delta=\Delta_c$ the critical value of the momentum vanishes
and becomes non-zero continuously for $\pi/2>\Delta > \Delta_c$. The
critical value $k_c(\Delta)$ determines the origin of the pitchfork
bifurcation. One of the branches of the pitchfork bifurcation moves
towards larger values of the frequency but the other towards smaller
values, eventually terminating at an end point which for the case of
the figure  (\ref{fig:bifurca}) is $k_e \sim 0.5 ~m_\nu$. For large
values of the momenta we find that the roots determine the same
dispersion relation as for the previous cases, which is given by
eqn. (\ref{largek}) and the spectral weight is also given by eqn.
(\ref{Zfast}).

While this  phenomenon of a pitchfork bifurcation in the spectrum is
a noteworthy aspect of neutrino collective excitations in the
medium, the narrow window in the parameter $\Delta$ within which
this phenomenon emerges is rather restricted and without a
particular significance within the standard model.

\subsubsection{$\Delta > \pi/2$.}

\begin{figure}[h!]
\begin{center}
\includegraphics[height=3in,width=4in,keepaspectratio=true]{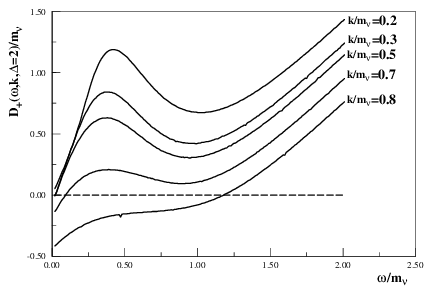}
\caption{Emergence of the gapless branch for
$D_+(\omega,k,\Delta)$ for $\Delta > \pi/2$.  }
\label{fig:dpgaples}
\end{center}
\end{figure}

For $\Delta > \pi/2$  only the gapless branch of
$D_+(\omega,k,\Delta)$ is available and the bifurcated spectrum
ends. This is depicted in fig. (\ref{fig:dpgaples}) which displays
the roots near the origin for small values of $k/m_\nu$. These
roots move continuously towards larger values of the frequency for
larger momenta.

\begin{figure}[h!]
\begin{center}
\includegraphics[height=3in,width=4in,keepaspectratio=true]{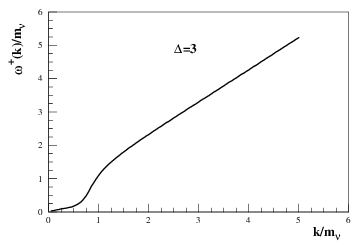}
\caption{Quasiparticle dispersion relation for  $\Delta=3$.}
\label{fig:disprelDG}
\end{center}
\end{figure}

 The dispersion relation for
$\Delta>\Delta_c$ is displayed in fig. (\ref{fig:disprelDG}) for
$\Delta=3$. For $k \ll m_\nu \Delta$ the dispersion relation is
linear but \emph{below the  light cone} and the group velocity
agrees with the result (\ref{gaplessvg}), whereas for $k \gg m_\nu
\Delta $ the dispersion relation is found to be

\be \omega^+(k) = k +
\frac{m^2_\nu}{k}\Big(1+\frac{1}{2\cos^2\theta_w}\Big)+\cdots
\label{dispGC}\ee

\noindent which is the same as for the large momentum limit of the
previous cases. Thus the spectrum interpolates continuously between
the soft non-perturbative region $k \ll m_\nu$ and the hard
perturbative region $k \gg m_\nu$ with a rapid crossover at $k
\approx m_\nu$.

The large momentum limit of the spectral weight in this case is also
given by eqn. (\ref{Zfast}).

\subsection{Quasi-hole spectrum.}\label{Dmin}

Just as in the previous section, we begin the analysis by revisiting
the case $\Delta=0$. In this case

\be D_-(\omega,k,0) =
m_\nu\Bigg\{\bw+\bk-\frac{C}{2\bk}\Big[\Big(1+\frac{\bw}{\bk}\Big)\ln\Big|
\frac{\bw+\bk}{\bw-\bk} \Big|-2\Big]\Bigg\}~~;~~C= 1+
\frac{1}{2\cos^2\theta_w} \label{DmD0}\ee

For $\bw = 0$ this function is positive and for $\bw \gg \bk$ it
has the asymptotic behavior $ D_-(\omega,k,0) \sim \omega$.
However for $\omega \sim k$ the logarithm gives rise to a sharp
downward spike and the function becomes negative near the light
cone for \emph{any} value of $k$. Therefore the function features
\emph{two roots} one above and one \emph{below} the light cone.
The dispersion relation for small $k$ is obtained from eqn.
(\ref{smaK}) for $\Delta=0$.

As $k \rightarrow 0$ the root above the light cone approaches the
gap (\ref{gapHTL}) from below with group velocity $c_-=-1/3$ and
the one below the light cone approaches the origin from above with
group velocity $c_- = 1/3$. This is precisely the gapless branch,
which is the solution of the gap equation for \emph{any} $\Delta$,
including $\Delta=0$ as mentioned in subsection (\ref{gaps}). This
gapless branch has not been discussed in previous studies of the
(HTL) approximation in unbroken gauge
theories\cite{weldon,pisarski,lebellac} because it is below the
light cone and is strongly Landau damped in those cases. We will
discuss below the width of collective excitations in this branch
(see section (\ref{damping})).

This simple analysis  combined with the linear dispersion relation
 for long wavelength collective modes in the gapless branch with
 \emph{positive energy} roots of $D_-(\omega,k,\Delta)$
   indicates that the gapless branch is a solution of
$D_-(\omega,k,\Delta)=0$ for $\Delta < \pi/2$. Positive energy
solutions along this branch correspond to gapless quasiholes. For
$\Delta \ll 1$ the function $D_-(\omega,k,\Delta)$ has a slope
$\propto 1/\Delta^2$ for both the gapless branch and the branch
with the smallest gap $\sim \Delta$. Therefore the spectral weight
for these branches is of order $\Delta^2 \ll 1$ and therefore
negligible with respect to that of the branch with gap $\sim m_\nu
(1+\frac{1}{2\cos^2\theta_w})^\frac{1}{2}$.

\begin{figure}[h!]
\begin{center}
\includegraphics[height=3in,width=4in,keepaspectratio=true]{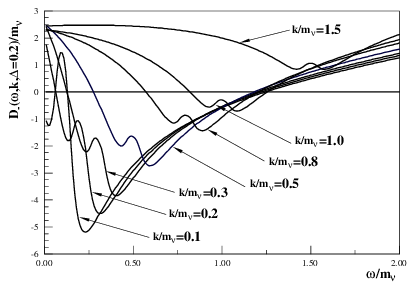}
\caption{Roots of $D_-(\omega,k,\Delta=0.2)/m_\nu$ for
$\frac{k}{m_\nu}=0.1,0.2,0.3,0.5,1.0,1.5$ respectively.}
\label{fig:rootsDmino2}
\end{center}
\end{figure}

For $\Delta=0$ the origin of the  two roots, above and below the
light cone, is traced to the logarithmic singularity at $\omega
=k$ in eqn. (\ref{DmD0}). However inspection of the functions $LP$
and $LM$ given by eqns. (\ref{LpW},\ref{LmW}) reveals that for any
$\Delta \neq 0$ this singularity is screened. For small $\Delta$ a
similar situation featuring two roots is expected for small $k$.
However for larger $\Delta$  and $k$ the screening of the light
cone singularity will not allow the function to become negative.
This feature is clearly displayed in fig. (\ref{fig:rootsDmino2})
for $\Delta=0.2$.

\begin{figure}[h!]
\begin{center}
\includegraphics[height=3in,width=4in,keepaspectratio=true]{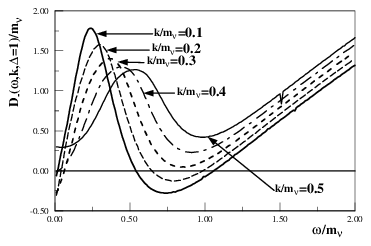}
\caption{Roots of $D_-(\omega,k,\Delta=1)/m_\nu$ for
$\frac{k}{m_\nu}=0.2,0.3,0.4,0.5$ respectively.}
\label{fig:rootsDmin}
\end{center}
\end{figure}

Inspection of this figure reveals the presence of \emph{three}
branches as anticipated, two gapped branches and one gapless
branch, which is manifest as the root closest to the origin for
small values of $k$ in the figure. The gapless branch terminates
at an end point value of the momentum $k_e$. Fig.
(\ref{fig:rootsDmino2}) shows that for larger values of $k$ there
are no available roots (see for example the curve for $k/m_\nu
=1.5$ in the figure). Therefore for larger values of the momentum
all the branches terminate at different end point values that
depend on $\Delta$. It is found numerically that this behavior is
consistent for all values of $\Delta < \Delta_c$: one gapless and
two gapped branches, each branch  terminates at a different end
point value of the momentum which depends on $\Delta$. In
particular the end point value for the gapless branch diminishes
continuously as $\Delta \rightarrow \pi/2$. Figure
(\ref{fig:rootsDmin}) displays the function
$D_-(\omega,k,\Delta=1)$ as a function of $\omega$ for several
values of $k$. Three branches are clearly displayed in this figure
as well as the termination of these branches at particular values
of the momentum $k$, for example only the gapless branch remains
for $k \gtrsim 0.3 m_\nu$ and finally no roots are available for
$k > 0.4 m_\nu$. The quasihole spectrum features end points for
all the branches and there are no propagating quasihole states for
$k \gg m_\nu$.

\begin{figure}[h!]
\begin{center}
\includegraphics[height=3in,width=3in,keepaspectratio=true]{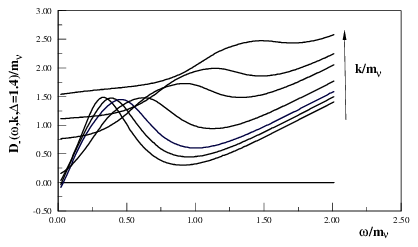}
\includegraphics[height=3in,width=3in,keepaspectratio=true]{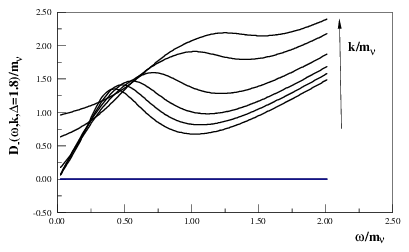}
\caption{Roots of $D_-(\omega,k,\Delta=1.4)/m_\nu$ and
$D_-(\omega,k,\Delta=1.8)/m_\nu$ for
$\frac{k}{m_\nu}=0.1,0.2,0.3,0.5,1,1.3$ respectively. The arrows
indicate increasing values of $k/m_\nu$} \label{fig:rootsDmin1418}
\end{center}
\end{figure}

Figure (\ref{fig:rootsDmin1418}) display two relevant cases. The
left panel shows the function $D_-(\omega,k,\Delta)$ for $\Delta_c
< \Delta < \pi/2$. This figure  shows that the gapped branches
have disappeared and there is a rather small window of (small)
momentum within which there are gapless quasihole states. Finally
the figure on the right corresponds to $\Delta > \pi/2$, which
clearly reveals that there are no available roots and all of the
quasihole branches have disappeared, including the gapless branch,
consistently with the previous discussion.

In summary, the spectrum of quasihole states features three
branches for $\Delta < \Delta_c$, one gapless and two gapped
branches, each branch terminates at particular end-point values of
the momentum which are functions of $\Delta$. For $\Delta_c <
\Delta < \pi/2$ the gapped branches disappear and only the gapless
branch remains but terminating at an end point value of the
momentum which is rather small. For $\Delta > \pi/2$ there are no
quasihole states. Only for  $\Delta \ll 1 $ there are  quasihole
states for $k \gg m_\nu$, this is consistent with the existence of
quasihole roots for arbitrarily large $k$ in the $\Delta=0$
case\cite{weldon,lebellac}.

For generic values of $\Delta$,  quasihole states are
non-perturbative, soft collective excitations available only for
momenta $k \lesssim m_\nu$.

The spectral weight of the quasihole excitations is related to the
slope of the function $D_-(\omega,k,\Delta)$ as a function of
$\omega$ by eqn.  (\ref{Zpm}). In the general case the spectral
weight must be found numerically, however, inspection of the figures
in this section clearly reveals that the spectral weight associated
with the gapless excitations is much \emph{smaller} than that of the
collective modes on the gapped branches. For $\omega \ll m_\nu$ the
slope of the function $D_-(\omega,k,\Delta)$ is $\approx 1/\Delta^2$
for $\Delta \ll 1$. Therefore gapless collective modes yield smaller
contributions to the spectral density than the gapped excitations.

\subsection{Damping rates of collective excitations:}\label{damping}

The damping rates or widths of the quasiparticle and quasihole
excitations are given by eqn. (\ref{widths}) \noindent where
$Z_\pm(k)$ are the residues given by eqn. (\ref{Zpm}), and the
imaginary parts are given by

\be \textrm{Im}D_{\pm}(\omega_\pm(k),k) =
\left[\textrm{Im}\sigma^0_W(k,\omega)\mp
\textrm{Im}\sigma^1_W(k,\omega)\right]+\left[\textrm{Im}\sigma^0_Z(k,\omega)\mp
\textrm{Im}\sigma^1_Z(k,\omega)\right]\label{imaparts}\ee

The imaginary parts
$\textrm{Im}\sigma^0_{W,Z}(k,\omega);~\textrm{Im}\sigma^1_{W,Z}(k,\omega)$
are given by the expressions (\ref{Imlisig0W}),\ref{Imlisig1W})
for the charged current interactions and similar expressions
obtained via the replacement (\ref{repla}) for the neutral current
interactions. An analysis of the delta functions in these
expressions determines the region of support of the imaginary
parts and thereby establish whether there is a non-trivial width
for the collective excitations with dispersion relations
$\omega_\pm(k)$.

\begin{figure}[h!]
\begin{center}
\includegraphics[height=3in,width=4in,keepaspectratio=true]{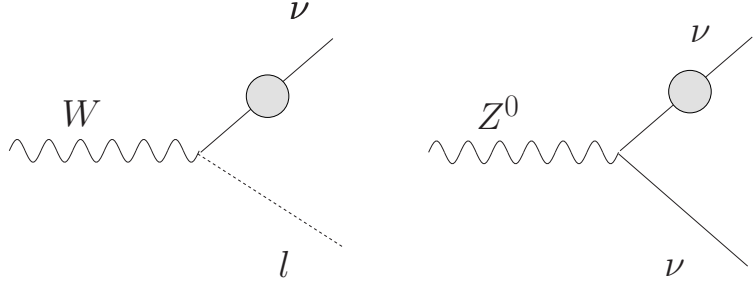}
\caption{Decay of vector bosons.  $W\rightarrow \nu+\overline{l}$
via the charged current or $Z \rightarrow \nu \overline{\nu}$ via
the neutral currents. The blob in the external neutrino line
corresponds to a quasiparticle or quasihole.  }
\label{vecbosdecay}
\end{center}
\end{figure}

\begin{itemize}

\item{{\bf{Region of support for $\delta(\omega-p-W_q)$:}} this
delta function corresponds to the ``decay'' process $\nu
\rightarrow W+\overline{l }$ for the charged current and $\nu
\rightarrow Z+ \overline{\nu}$ for the neutral current. Clearly
the neutral current contribution cannot be fulfilled. For the
charged current contribution, the quasiparticle and quasihole
\emph{could} decay provided $m_\nu > M_W(T)$. In principle the
kinematics for this process \emph{could} be satisfied since near
the ``transition'' the mass of the vector boson may be small.
However  this would entail that

\be g T > M_W(T)  \label{ineq} \ee

\noindent which, however, contradicts the bounds
(\ref{cubicbound}), (\ref{tightbound}) which determine the domain
of validity of the perturbative expansion, the latter one being
the most conservative. Therefore within the domain of reliability
of the perturbative expansion invoked in this study, the
kinematics resulting from this delta function cannot be fulfilled
on the quasiparticle or quasihole mass shell. The delta function
$\delta(\omega+W_q+p)$ has support only for $\omega <0$ and
corresponds to the decay of a negative energy neutrino.  }

\item{{\bf{Region of support for $\delta(\omega+p-W_q)$:}} this
delta function corresponds to the decay process $W \rightarrow
\nu+\overline{l}$ (the term with $\delta(\omega-p+W_q)$
corresponds to the decay into antineutrino-lepton) and it has
support on the neutrino mass shell for $M_W(T) > m_\nu$ for a
massless lepton, or $Z\rightarrow \nu + \nu$ for $M_Z(T) > 2
m_\nu$. Either of the bounds (\ref{cubicbound}),
(\ref{tightbound}) which determine the domain of validity of the
perturbative expansion guarantees that these processes are
kinematically allowed in the region of validity. Therefore this
analysis leads to the conclusion that the \emph{decay} of the
vector bosons into neutrino-lepton pairs (charged current) or
neutrino pairs (neutral current) lead to a \emph{width} for the
neutrino collective excitations. These processes are depicted in
fig. (\ref{vecbosdecay}). }

\end{itemize}

The fact that the decay of vector bosons into neutrinos leads to a
width for the neutrino states can be easily understood from a simple
kinetic argument. Consider the kinetic equation for the time
evolution of the neutrino distribution function. This equation is
obtained by a gain minus loss argument as follows. Consider first
the charged current interaction: W-decay, $W \rightarrow \nu
~\overline{l}$, leads to a gain in the neutrino distribution and the
recombination $\nu+\overline{l} \rightarrow W$ leads to a loss. The
gain and loss terms are obtained from the transition probability per
unit time for each process leading to the following kinetic equation

\be \frac{d n_{\nu}(k)}{dt} = \int \dbarq
\frac{|\mathcal{M}_W(q)|^2}{4 W_q \,p}
\Bigg[(1-n_\nu(k))(1-n_l(p))N_W(W_q)-n_\nu(k)\,n_l(p)\,(1+N_W(W_q))\Bigg]\delta(W_q-p-\omega(k))\label{kineq}
\ee

\noindent where $|\mathcal{M}_W(q)|^2$ is the transition probability
matrix element, $\vp=\vk-\vq$ and $n_l\,,\,N_W$ are the
 distribution functions for leptons and  for vector bosons respectively. A similar
contribution arises for the neutral currents which can be simply
obtained from the expression above by the corresponding
replacements and $n_l\rightarrow n_\nu$. The linearized kinetic
equation around the equilibrium distributions, is obtained by
writing $n_{\nu}(k)=N_f(k)+\delta~n_\nu(k)$ with $N_f(k)$ the
Fermi-Dirac distribution function and keeping the lepton and
vector boson distribution functions to be those of equilibrium,
namely $n_l(p)=N_f(p)~;~N_W(W_q)=N_b(W_q)$. Keeping only linear
terms in $\delta~n_\nu(k)$ one obtains the kinetic equation in the
relaxation time approximation

\be \frac{d~\delta n_{\nu}(k)}{dt} = - \delta n_{\nu}(k) \int
\dbarq \frac{|\mathcal{M}_W(q)|^2}{4 W_q \,p}
\Bigg[N_b(W_q)+N_f(p)\Bigg]\delta(W_q-p-\omega(k)) \,.\ee

Thus we see that the decay rate of the distribution function

\be \gamma(k) = \int \dbarq \,\frac{|\mathcal{M}_W(q)|^2}{4 W_q
\,p}\, \Bigg[N_b(W_q)+N_f(p)\Bigg]\delta(W_q-p-\omega(k))\ee

\noindent is  identified with last terms in the imaginary parts
(\ref{Imlisig0W}),\ref{Imlisig1W}). The damping rate for
quasiparticles $\Gamma(k)$ is simply related to the relaxation
rate of the distribution function as $\gamma(k)=2\,\Gamma(k)$
\cite{weldon,lebellac,nosfermions,tututi}. The fact that the decay
of a \emph{heavier} particle implies a width for the light
collective modes in a thermal bath has already been recognized in
refs.\cite{nosfermions,man}.

A simple analysis of the region of support of $
\delta(W_q-p-\omega)$ in (\ref{Imlisig0W}),\ref{Imlisig1W}) shows
that for $\omega$ above the light cone ($\omega > k$)  the loop
momentum integral is restricted to the range $q^+\leq q \leq q^-$
where

\be q^+ = \Bigg|\frac{M^2_W(T)-(\omega+k)^2}{2(\omega+k)}
\Bigg|~~;~~ q^- = \frac{M^2_W(T)-(\omega-k)^2}{2(\omega-k)}\ee

We find that the charged current contribution to the imaginary
parts are given by

\be  \textrm{Im}D^W_{+}(\omega_+(k),k) = \frac{g^2}{32\pi\,k}
\int_{q^-(\omega_+(k))}^{q^+(\omega_+(k))} \frac{q \, dq}{W_q}
\Big(Q_{0}(\vec{p},-\vec{q})- \widehat{\bf
k}\cdot\vec{Q}(\vec{p},-\vec{q}) )\Big)\Big[
N_f(p)+N_b(W_q)\Big]\label{ImDplus}\ee

\noindent with

\be p=W_q-\omega_+(k)~~;~~ \widehat{\bf q} \cdot \widehat{\bf k}=
\frac{q^2-k^2-p^2}{2qk}\label{ImDplusi} \ee

\noindent and

 \be  \textrm{Im}D^W_{-}(\omega_-(k),k) = \frac{g^2}{32\pi\,k}
\int_{q^-(\omega_-(k))}^{q^+(\omega_-(k))} \frac{q\, dq}{W_q}
\Big(Q_{0}(\vec{p},-\vec{q})+ \widehat{\bf
k}\cdot\vec{Q}(\vec{p},-\vec{q})\Big)\Big[ N_f(p)+N_b(W_q)\Big]
\label{ImDmin} \ee

\noindent with

\be p=W_q-\omega_-(k)~~;~~\widehat{\bf q} \cdot \widehat{\bf k}=
\frac{q^2-k^2-p^2}{2qk}\label{ImDmini} \ee

The contribution from neutral current interactions is found from
the expressions above by the replacement (\ref{repla}). The
damping rate can be obtained numerically in the general case, but
analytic progress can be made in two limits: quasiparticle
(quasihole) at rest, namely $k=0$, and for fast moving
quasiparticles  $T\gg k \gg m_{\nu}$. As discussed in the previous
section, only for $\Delta \ll 1$ are there quasihole states for $k
\gg m_\nu$.

Furthermore within the domain of validity of the perturbative
expansion determined  the tighter bound (\ref{tightbound}) it
follows  that $M_W(T)/m_\nu \gg 1$ which results in several
simplifications.

\vspace{2mm}

{\bf Quasiparticle and quasihole at rest:} for $k=0$ the damping
rates of quasiparticle and quasiholes are the same. It can be
obtained directly from the expressions for the imaginary parts
(\ref{Imlisig0W}),\ref{Imlisig1W}) by recognizing that for $k=0$ the
contribution from (\ref{Imlisig1W}) must vanish by rotational
invariance or alternatively by computing the integral in the
expressions above in the limit $q^+\rightarrow q^-$. The final
result for the damping rate is

\be \label{restwidth}\Gamma_\pm(\omega_g) =
\frac{g^2\,\omega_g}{32\pi}Z[\omega_g]\left[F[M_W(T)]+\frac{F[M_Z(T)]}{2\cos^2\theta_w}\right]\ee

\noindent with

\bea F[M]  & = & \left[
\frac{M^2-\omega^2_g}{\omega^2_g}\right]^2\left[1+\frac{\omega^2_g}{2M^2}\right]\left[N_f(q^*)+N_b(W(q^*))\right]
\label{funF}\\
q^* & = &
\frac{M^2-\omega^2_g}{2\omega_g}~~;~~W(q^*)=\frac{M^2+\omega^2_g}{2\omega_g}
\label{qstar} \eea

\noindent where $\omega_g$ is a non-zero solution of the gap
equation (\ref{gapeqn}) and $Z[\omega_g]$ is the residue
corresponding to zero momentum quasiparticles or quasiholes.

Since $\omega_g \leq m_\nu$ and in the domain of validity of the
results $M_{W,Z}(T) \gg m_\nu$,  the expression for the function
$F[M]$ in eqn. (\ref{funF}) simplifies to

\be F[M] \sim \left[ \frac{M}{\omega_g}\right]^4
~\frac{1}{\sinh\big[\frac{M^2}{2\,\omega_g T} \big]} \ee

Therefore the width for quasiparticles and quasiholes at rest is
given by

\be \label{restwidthfin}\Gamma_\pm(\omega_g) =
\frac{2~m_\nu}{\pi}Z[\omega_g]\left[
\frac{m_\nu}{\omega_g}\right]^3~\Delta^2
\Bigg\{\frac{1}{\sinh\big[\Delta \frac{m_\nu}{\omega_g} \big]} +
\frac{1}{2\cos^6\theta_w}
\frac{1}{\sinh\big[\frac{\Delta}{\cos^2\theta_w}
\frac{m_\nu}{\omega_g} \big]}\Bigg\}\ee

For $\Delta \ll 1$ the gap for the lowest gapped branch is $\omega_<
\sim m_\nu~\Delta$ and for the upper branch is $\omega_> \sim m_\nu$
while the residues are $Z[\omega_g]\sim \Delta^2 $ for the lower gap
and $Z[\omega_g] \sim 1/2$ for the upper branch.

Hence  it follows that for $\Delta \ll 1$

\bea \Gamma(\omega_<)  & \approx &  \omega_<  \label{gamlow}\\
\Gamma(\omega_>)  & \approx &  \Delta ~\omega_> \label{gamup}\eea

Therefore for $\Delta \ll1$ the lowest branch is strongly damped
while the upper branch is weakly damped.

For $\Delta \sim 1~~;~~\Delta < \Delta_c$ both the lower and upper
branch have gaps of order $m_\nu$ (up to factors of order one, see
for example fig. (\ref{fig:rootsDplus})) and $Z[\omega_g]\sim 1$,.
Hence for $\Delta < \Delta_c$ but $\Delta \sim 1$,

\be \Gamma(\omega_g) \sim \omega_g \label{gammadel1}\ee

\noindent and collective excitations at rest on either branch are
strongly damped.

{\bf Fast moving quasiparticles :} The other limit that is relevant
for a comparison with the perturbative results and tractable
analytically is that of $\Delta \gtrsim 1$ and fast moving
quasiparticles with $ M(T) \gg k \gg m_\nu \,\Delta$. In this limit
only quasiparticles are available since the spectrum of quasiholes
terminates at an end point of order $k_e \sim m_\nu \Delta$  for not
too small $\Delta$ (see figs.
(\ref{fig:rootsDmino2}-\ref{fig:rootsDmin1418})). For $k \gg
m_\nu\Delta$ the quasiparticle dispersion relation is near the light
cone and given by eqn. (\ref{largek}) and the spectrum of
quasiparticles is perturbatively close to the free field spectrum,
therefore a comparison to the perturbative results is meaningful.

In this limit $q_+/T \gg 1$  and $q_- \gg M(T)$ the upper limit of
the integral can be taken to $q_+\rightarrow \infty$.  The remaining
integrals are now straightforward and after combining the results
from charged and neutral currents we find

\be \Gamma_+(k) \sim \frac{ g^2\,T}
{32\pi}\frac{M^2_{W}(T)}{k^2}\ln\Big[\frac{4k}{m_\nu
\Delta}\Big]\Big(1+\frac{1}{2\cos^4\theta_w}
\Big)\label{gamapfast}\ee

Thus in the limit $k \gg m_\nu \Delta$ quasiparticles are weakly
damped since

\be \Gamma_+(k)/k \sim \Delta \frac{m^3_\nu}{k^3}
\ln\Big[\frac{4k}{m_\nu \Delta}\Big] \ll 1\ee

\section{Domain of validity of the HTL approximation: possible
caveats.}\label{domain}

In the (HTL) approximation we have used the bare propagators for the
internal fermion and vector boson lines. Since the (HTL)
approximation relies on loop momentum $q \sim T$ the region of
validity of the (HTL) approximation is determined by the region in
which the self-energy corrections to the \emph{internal} lines are
perturbative with respect to the loop momentum scale. The study of
the previous section  indicates that vector boson exchange will lead
to a radiative correction to the internal fermionic line yielding a
mass scale $m_\nu \sim gT$. Thus the self energy correction to the
lepton or neutrino internal lines is indeed perturbative small in
weak coupling.

\begin{figure}[h!]
\begin{center}
\includegraphics[height=3in,width=4in,keepaspectratio=true]{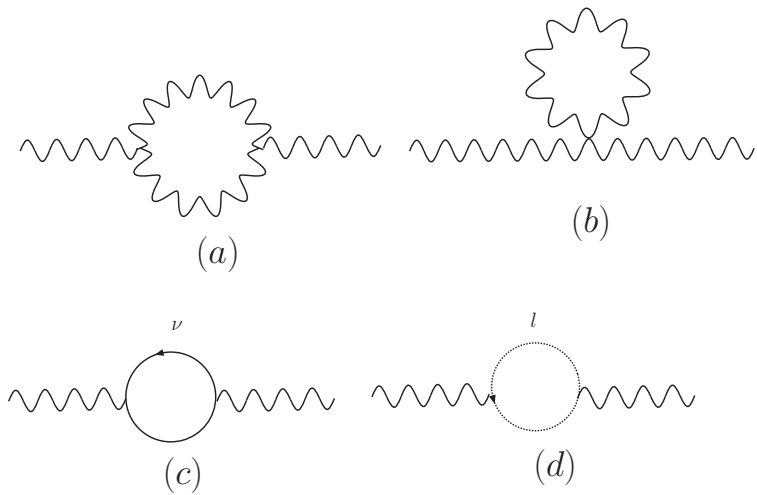}
\caption{One loop diagrams contributing to the self-energy of
charged and neutral vector bosons.  } \label{vecbosSE}
\end{center}
\end{figure}

Furthermore the vector bosons themselves acquire self energy
corrections through fermion loops \emph{and} vector boson loops
through the non-abelian self-interaction. These are displayed in
fig. (\ref{vecbosSE}).

The self-energy diagrams with fermion loops (diagrams (c) and (d)
as well as similar diagrams with quarks for the neutral vector
boson) yield a typical (HTL) self-energy correction of order $g^2
T^2$\cite{pisarski,lebellac} and are perturbative for external
momentum $q \gg gT$ or for soft external momentum if $M(T)\gg gT$,
where $M(T)$ stands generally for the vector boson mass.

These self energy corrections yield a Debye (electric) mass $m_D
\sim g T$\cite{pisarski,lebellac,manuel}. In non-abelian theories, a
magnetic mass $m_m \sim g^2 T$ is also generated at high
temperature\cite{pisarski,lebellac}.

Since we are considering the high temperature limit with $T\gg
M_{W,Z}(T)$ these contributions are perturbative for soft external
momentum in the regime

\be T \gg M(T)  \gg g T\label{bound1}\ee

A tighter bound emerges from consideration of the self-energy
contribution from non-abelian loops, such as (a) and (b) in
fig.(\ref{vecbosSE}). The (HTL) power
counting\cite{pisarski,lebellac} for these diagrams is modified by
the large momentum behavior of the propagators in the (physical)
unitary gauge. Since for large momentum the vector boson propagators
$G_{\mu\nu}(q) \sim q_\mu q_\nu/q^2 M^2(T) \sim 1/M^2(T) $ the
\emph{naive} (HTL) counting assigns a factor of $1/M^2(T)$ for each
vector boson line, rather than the usual
$1/T^2$\cite{pisarski,lebellac}.

Therefore, according to this modified (HTL) power counting the
vector boson loop with cubic (momentum dependent) vertex in diagram
(a) in fig. (\ref{vecbosSE}), would yield a self energy contribution

\be \Sigma_a \propto \frac{g^2 T^6}{M^4(T)} \label{cubicSE}\ee

\noindent Four powers of temperature from the loop with two extra
powers from the momentum dependence of each vertex and four powers
of $M$ in the denominator because each vector boson line yields a
factor $1/M^2(T)$. Similarly the vector boson loop with quartic
(momentum independent) vertex in diagram (b) in fig.
(\ref{vecbosSE}),  would yield a self energy contribution

\be \Sigma_b \propto \frac{g^2 T^4}{M^2(T)} \label{quarticSE}\ee

It is clear that \emph{if} the usual (HTL) power counting holds for
these diagrams  as $M(T) \rightarrow 0$ the infrared divergences
lead to a breakdown of the perturbative expansion, requiring a
resummation.

 However for sufficiently
high temperatures but \emph{ below} the symmetry restoration scale
$T_{EW}$, because the standard model features a smooth crossover
from the broken to the unbroken ``phase'',  there is a regime for
which $T_{EW}> T > M(T)$ within which a perturbative treatment is
reliable.

Within this window of validity of perturbation theory, the validity
of the (HTL) approximation will be further bound by the region in
which the self-energy contribution to the gauge vector bosons is
perturbatively small. The analysis above reveals that for $T \gg
M(T)$ diagram (a) in fig.(\ref{vecbosSE}) gives the largest
contribution, since at high temperature the loop momentum is $q\sim
T$ the self energy contribution (\ref{cubicSE}) will be perturbative
if

\be T > \frac{g T^3}{M^2(T)} \Longrightarrow M(T) > g^{\frac{1}{2}}
T \label{cubicbound}\ee

A tighter and more conservative bound is obtained from the
observation that in using the free field theory propagators for the
internal lines the dispersion relation of on-shell vector bosons has
been used. In particular the polarization vectors involve the bare
mass $\sim M(T)$. Hence a tighter bound is obtained by requiring

\be M(T) > \frac{g T^3}{M^2(T)} \Longrightarrow M(T) >
g^{\frac{1}{3}} T \label{tightbound}\ee

The bounds above are the result of assuming the validity of the
\emph{naive} argument that combines the  (HTL) power counting with
the large momentum behavior of the vector boson propagator.

However this simple counting argument ignores the subtleties
associated with the underlying gauge invariance as well as
potential cancellations. It is \emph{plausible} that the term of
$\mathcal{O}(g^2 T^6/M^4(T)$ vanishes identically and that there
is a cancellation of the terms of $\mathcal{O}(g^2 T^4/M^2(T)$
between the diagrams (a) and (b) as a result of Ward identities.
\emph{If } such is the case, then only a contribution of
$\mathcal{O}(g^2 T^2)$ would remain in which case the perturbative
expansion is reliable for

\be M(T) \gg g  T  \label{lobound}\ee

An indication  that these cancellations are \emph{plausible} is
found  precisely in  the neutrino self-energy in the high
temperature limit obtained in the appendix and section
(\ref{selfenergy}). The \emph{naive} (HTL) power counting combined
with the large momentum behavior of the vector boson propagator
$\propto 1/M^2(T)$ would lead to the estimate $\sim g^2 T^3/M^2(T)$
for the first two terms and $g^2 T^4 / M^2 k$ for the last two terms
in the self-energy corrections (\ref{sig0W},\ref{sig1W})). However,
remarkable cancellations in both terms lead to $\sim g^2 T^2
\omega/M^2(T)$ for the first two terms and $g^2 T^2/k$ for the last
two terms, and perturbation theory is reliable for $M^2(T)\gg gT$.
Of course, whether these cancellations also occur in the self-energy
of the vector bosons must be studied in detail.

In either of these cases (\ref{cubicbound}-\ref{lobound}) the vector
boson masses from symmetry breaking $M(T)$ are much greater than the
Debye  or magnetic masses, $m_D \sim gT$, $m_m\sim g^2 T$
respectively.

Even if the more conservative bound (\ref{tightbound}) survives
deeper scrutiny, within \emph{strict} perturbation theory a regime
in which the (HTL) approximation is reliable within this bound
clearly exists, namely $T > M(T)> g^{\frac{1}{3}} T$. However, this
analysis warrants a deeper study of the  screening corrections to
the vector bosons to provide a reliable bound for the validity of
the perturbative expansion at high temperature.

While the (HTL) approximation is well
understood\cite{pisarski,lebellac} in the \emph{absence} of symmetry
breaking, the interplay between the generation of a Debye and
magnetic masses and the mass from symmetry breaking $M(T)$ below the
critical temperature is not yet completely understood.

In ref.\cite{manuel} the Debye screening corrections to the vector
boson masses was studied in the temperature regime $M(T) \ll T \ll
\sqrt{12}\, v(0)$, where $v(0)$ is the expectation value of the
neutral component of the Higgs doublet \emph{at zero temperature}.
At finite temperature it is found in this reference that

\be v(T)= v(0)\left(1- \frac{T^2}{12 \, v^2(0)}\right) \ee

Therefore the regime of validity of the (HTL) approximation as
stated in ref.\cite{manuel} corresponds to temperatures \emph{much
smaller} than the critical.

In this regime of temperature it was found in ref.\cite{manuel} that
the vector bosons feature a \emph{longitudinal} squared mass of the
form

\be M^2_l(T)= M^2(T) + g^2 T^2 \mathcal{A}(\theta_w)
\label{longmas}\ee

\noindent where $\mathcal{A}(\theta_w) $ is a simple function of the
Weinberg angle, and a \emph{transverse} mass squared

\be M^2_t(T) = M^2(T) \label{tranmas}\ee

A magnetic mass term is expected at order
$g^4$\cite{pisarski,lebellac}. These results are valid well below
the critical temperature, and to the best of our knowledge there is
as yet no  clear understanding of the validity of the (HTL)
approximation \emph{near} the critical temperature.
Ref.\cite{manuel} concludes by stating that a resummation program
must be implemented, a statement that becomes more relevant near the
critical temperature. This discussion is meant to bring to the fore
the relevant but largely unexplored question of the validity of the
(HTL) near the critical temperature.

Such a study is obviously beyond the realm of this article and is
deferred to future investigations.

\section{Discussion of the results}\label{disc}

\begin{itemize}

\item{{\bf{Collective modes:}} There are several  remarkable
  differences between the spectrum of collective modes obtained
  above and those for fermionic collective excitations in QCD
   or QED\cite{klimov,weldon,pisarski,lebellac}. To begin with, in the region in which
   perturbation theory is reliable, $T\gg M_{W,Z}(T)\gg gT$, the
   spectrum depends on the mass parameter $m_\nu= gT/4$, which determines the
   \emph{chirally symmetric gaps} in the spectra, as well as
   the   dimensionless ratio $\Delta=M^2_W(T)/2m_\nu T$ and features
   in general several branches. Gapped branches for quasiparticles
   and quasiholes are present for $\Delta < \Delta_c \sim
   1.275\cdots$ a gapless branch for quasiholes exists for $\Delta <
   \pi/2$ which becomes a gapless branch of quasiparticles for
   $\Delta > \pi/2$. The quasihole branches terminate at end points that depend on
   the value of $\Delta$ and for large momentum $k\gg m_\nu \Delta$
   there are \emph{no quasihole branches} available.  For $\Delta_c < \Delta <\pi/2$ the
   quasiparticle spectrum features a pitchfork bifurcation with two
   branches emerging, the one with decreasing frequency terminates
   at an end point but the other with increasing frequency continues
   and merges asymptotically with the free field dispersion
   relation. For $\Delta > \pi/2$ the collective modes are gapless quasiparticles whose
   dispersion relation lies below the light cone for $k \lesssim m_\nu$ and approaches
    the free field    dispersion relation for $k \gg m_\nu$.  }

\item{{\bf{Gauge invariance:}} An important aspect that must be
discussed is the issue of gauge invariance. We have obtained the
neutrino self-energy in the unitary gauge. This is a physical gauge
in the sense that it displays only the physical excitations. In any
other covariant gauge there are unphysical degrees of freedom, in
particular unphysical Goldstone bosons with a Yukawa coupling to
neutrinos. In ref.\cite{dolivoDR} the neutrino self energy (at
temperatures much smaller than the mass of the vector boson) was
obtained in general covariant gauges an it is shown explicitly that
the dispersion relation is independent of the gauge parameter.
Furthermore, as discussed explicitly in
references\cite{weldon,pisarski,lebellac} the HTL approximation is
gauge invariant. While this body of work clearly points out that the
results obtained here are gauge invariant, there is a rather simple
argument that makes the gauge invariance manifest up to the one loop
order considered here. In the unitary gauge the only singularities
in the Feynman propagators correspond to the on-shell propagation of
\emph{physical} degrees of freedom and the unitarity of the S-matrix
follows directly from the Cutkosky rules\cite{peskin}. In particular
the imaginary part of the one-loop self-energy can be obtained
directly from the Cutkosky rules at tree level, and these only
involve the propagation of physical degrees of freedom. Since the
full one-loop self-energy is obtained from a dispersion relation,
the gauge invariance of the imaginary part guarantees the gauge
invariance of the self-energy up to the order considered. For
example, the calculation of decay rates in the standard model in the
Born approximation in the unitary gauge yield the physical
result\cite{peskin,ramond}, and the imaginary part of one-loop
diagrams is obtained from the Born transition elements by the
unitarity of the S-matrix, which is manifest in the unitary
gauge\cite{peskin} (in fact this \emph{is} the bonus and defining
property of the unitary gauge). Therefore we conclude that the
results obtained here are manifestly gauge invariant.}

 \item{{\bf{Width of collective excitations from vector boson decay:}}
  Another remarkable aspect that must be highlighted
 is that the width of the neutrino excitations in the medium is a result of the \emph{decay} of
 the vector bosons into neutrino or neutrino-lepton pairs. This novel mechanism is rather different
 from the usual collisional   relaxation that leads to a width of the quasiparticles. In particular collisional
 relaxation results in a width of order $G^2_F$ since the Born amplitude for such process must necessarily
 involve the exchange of a vector boson and is therefore is of order $G_F$. In contrast, the width acquired
 via the decay of a more massive vector boson is of order $G_F$
since the Born amplitude is of order $\sqrt{G_F}$. We emphasize that
this mechanism is present even if the vector boson population in the
medium is negligible, as can be gleaned from the fact that the term
from the  imaginary part of the self energy with support on the mass
shell of the neutrino excitation involves $N_f+N_b$. Therefore even
if $N_b \sim 0$ there will be a width for the neutrinos if there are
fermions (either neutrinos or leptons) in the medium. There is a
 simple interpretation of this rather striking result, which can be
 gleaned from the discussion of the width in terms of the kinetic
 equation (\ref{kineq}). While the first  term in
 (\ref{kineq}) which describes the gain in neutrino population from the
 decay of the vector boson, vanishes when $N_W\sim 0$, the second
 term does \emph{not} vanish when $N_W\sim 0$. This second term
  describes the \emph{recombination} process in which the
 neutrino in the medium annihilates with either another neutrino
 (neutral current) or a lepton (charged current) in the medium to
 produce a vector boson, the \emph{stimulated emission} makes this
 term non-vanishing even for vanishing population of vector bosons.
 Therefore this contribution to the width of neutrino excitations
 will \emph{always} be available in the medium provided the
 kinematics for energy-momentum conservation is fulfilled. Indeed
 this is the major restriction for the width that arises from this
 process, in order for the recombination process to be available,
 energy-momentum conservation require that the momentum of the
 lepton (or neutrino) in the bath must be very large (when the mass
 of the vector boson is much larger than that of the neutrino
 quasiparticle or quasihole) and therefore this process will probe
 the tail of the fermionic distribution function with the ensuing
 exponential suppression at low temperatures. This novel
 relaxational mechanism has been previously studied in different
 contexts in refs.\cite{nosfermions,man}.
 }
 \item{{\bf{Comparison with collisional damping:}} In
 references\cite{notzold,tututi} the \emph{collisional} damping
 rate of neutrinos was computed up to two loops order for temperatures much smaller than the vector boson
 mass.While we cannot directly compare those results to ours for
 two reasons: a) our study focuses on temperatures much larger
 than the (temperature dependent) vector boson mass, b) we have
 obtained the damping rate of quasiparticles and quasiholes rather
 than weakly interacting neutrino states, we can compare the
 results to obtain at least an estimate of the new effects. The
 result for the collisional damping rate obtained in
 refs.\cite{notzold,tututi} is

 \be \Gamma_c(\omega) \sim G^2_F T^4 \omega \label{gammacol}\ee

\noindent where $\omega$ is the energy of the neutrino, $\omega \sim
k$ for a fast moving neutrino in the medium. In ref.\cite{thomson} a
similar result for the collisional width was obtained but
temperature $T$ replaces the neutrino energy in eqn.
(\ref{gammacol}). These results are obtained from a kinetic equation
that includes the different scattering contributions, or
alternatively as in ref.\cite{tututi} from an analysis of the
imaginary part of the neutrino self-energy up to two loops. These
calculations take the external neutrino to be described by a free
field state and are restricted to temperatures below the vector
boson mass. Since fast moving quasiparticles are very similar to the
weakly interacting neutrino states, we can proceed to compare the
results for the width of the quasiparticle excitations for large
momentum $k \gg m_\nu\,\Delta$ given by eqn. (\ref{gamapfast}) with
the result of eqn. (\ref{gammacol}) for $\omega \sim k$. The ratio
of these results is

\be\label{ratiodamp} \frac{\Gamma_+(k)}{\Gamma_c(k)} \sim
\frac{1}{4\pi g^2} \Big[\frac{m_\nu \Delta}{k}\Big]^3
\ln\Big[\frac{4k}{m_\nu\Delta} \Big]\ee

In the expression above we have (unjustifiably) taken the vector
boson mass in (\ref{gammacol}) to be equal to $M(T)$ in our
calculation, with the purpose of offering a comparison. Granting
these caveats, it is clear that under the perturbative assumption
$M(T)\gg m_\nu \sim gT$ the ratio $\Gamma_+(k)/\Gamma_c(k) \gtrsim
1$ for $1 \gg \frac{m_\nu\Delta}{k} \gtrsim g^\frac{2}{3} $. Hence,
for a fairly wide range of neutrino energies, the relaxation
mechanism via the decay of vector bosons could be comparable to that
via collisions. While this comparison must be considered only an
estimate (given the caveats mentioned above), it at least
\emph{suggests} that the damping mechanism found in this article is
competitive with if not larger than the usual collisional relaxation
and must be included in treatments of non-equilibrium phenomena of
neutrinos in the early Universe.
 }

\item{{\bf  Regime of validity:} Although the results of ref.\cite{manuel}
are valid well below the critical temperature, they are encouraging
in that the vector bosons acquire a Debye screening mass of order $g
T$ which, \emph{if taken at face value near the critical
temperature} would entail that $\Delta \sim g \ll 1$. In this case
the dispersion relation of the collective modes will feature the
wealth of novel phenomena studied above. \emph{If} further studies
confirm that perturbation theory is reliable very near the critical
temperature and the vector boson acquires a screening mass of order
$g T$  then this would  translate  in that $\Delta \sim  g$.
Therefore, at least within perturbation theory there is a wide range
$1 > \Delta > g$ for which the collective modes feature the most
striking and richer aspects of the dispersion relations found above.
A similar conclusion is reached if it is required that $M^2(T) > g^4
T^2$ so that the transverse mass is larger than the magnetic mass to
avoid the breakdown of perturbation theory at long wavelengths.  If,
on the other hand, the more stringent constraints
(\ref{cubicbound},\ref{tightbound}) are shown to be the relevant
ones because of infrared phenomena associated with the magnetic mass
very near the critical temperature, then $\Delta \gtrsim 1$ and the
properties of the collective modes depart perturbatively from those
of the vacuum states. It is clear that a further assessment requires
a deeper understanding of the validity of the (HTL) program near the
critical temperature. }

\end{itemize}

\section{Conclusions}\label{conc}

Motivated to explore non-equilibrium properties of neutrinos that
could impact on thermal leptogenesis, or more generally in neutrino
transport in the early Universe,  we studied  the collective
excitations of neutrinos in the standard model at high temperature
but below the symmetry breaking scale $T_{EW}$. The main assumption
is  that the transition from the broken to the unbroken symmetry
states in the standard model is either a smooth crossover, as
supported by the lattice data with the current bound on the Higgs
mass, or of second order. In this scenario the expectation value of
the neutral scalar in the standard model vanishes continuously near
the transition resulting in that the mass of the vector bosons
becomes smaller, which in turn leads to  a large population of
vector bosons in the thermal bath. We have obtained the spectrum of
collective excitations for standard model neutrinos in the (HTL)
approximation in the regime $T\gg M_{W,Z}(T)\gg gT$ within which
perturbation theory is reliable. The excitation spectrum consists of
left handed positive energy negative helicity quasiparticles and
left handed positive energy and positive helicity quasiholes.
Antiquasiparticles and antiquasiholes carry negative energy and
opposite helicity assignments.  The excitation spectrum features a
chirally non-breaking mass scale

\be m_{\nu}  =  \frac{gT}{4} . \ee

\noindent and depends on the dimensionless ratio $\Delta= M^2_W(T)/2
m_\nu T$. For $\Delta < \Delta_c=1.275\cdots$, the quasiparticle
spectrum features two gapped branches  and the quasi-hole spectrum
features one gapless and two gapped branches of collective modes.

The lower quasiparticle branch as well as \emph{all} the quasihole
branches terminate at particular end point values of the momentum
that depend on $\Delta$. For $\Delta >\pi/2$ only a gapless
quasiparticle branch is available, with a dispersion relation that
is below the light cone for $k \ll m_\nu$ and asymptotically reaches
the free particle dispersion relation for $k \gg m_\nu$.

A novel feature  revealed by this study is that the decay of the
vector bosons into neutrinos and leptons leads to a damping rate for
the neutrino collective modes. A simple kinetic interpretation of
this phenomenon was given and analytic expressions for the damping
rates were obtained in the limit of small and large momentum of the
excitations. These are given by equations (\ref{restwidth}) and
(\ref{gamapfast}) respectively. For $\Delta \ll 1$ quasiparticles
and quasiholes at rest in the lower branch are heavily damped while
those in the upper branch are weakly damped, while for $\Delta
\gtrsim 1$ all collective modes \emph{at rest} are heavily damped.
Fast moving quasiparticles are weakly damped.

We have compared the damping rate of fast moving quasiparticles to
the results for the collisional damping rate of
refs.\cite{notzold,tututi,thomson} which is a two loop result
($\mathcal{O}(G^2_F)$), the ratio is given by equation
(\ref{ratiodamp}). The damping rate resulting from vector boson
decay at one loop can be \emph{larger} than the collisional
relaxation rate in a wide range of neutrino energy.

We have discussed in detail the gauge invariance of the results and
assessed the region of validity of the (HTL) approximation which
relies on a strict perturbative expansion.

The  novel aspects of the neutrino collective modes near the
crossover (or second order) transition to the unbroken phase as well
as their non-equilibrium relaxational dynamics studied in this
article could be important for a reliable assessment of mechanisms
for thermal leptogenesis\cite{fukugita,yanagida,buch,giudice}.

\vspace{2mm}

{\bf Further questions:} Clearly an aspect that warrants further
study is a more quantitative assessment of the screening corrections
to the gauge boson propagators, namely the diagrams displayed in
fig. (\ref{vecbosSE}), in particular diagrams (a,b). A detailed
evaluation of these self-energy corrections will yield a better
assessment of the domain of validity of the (HTL) approximation near
the critical temperature. At temperatures closer to the transition
(or crossover) a resummation of the perturbative series must be
invoked in order to correctly treat the infrared divergences in the
vector boson propagators\cite{breakdown}.

While we have focused on temperatures below the electroweak scale,
the study of neutrino collective excitations above this scale is
also inherently important for a deeper understanding of early
Universe neutrino cosmology. Above this scale, in the phase where
the $SU(2)\otimes U(1)$ is restored, new interaction vertices
between neutrino, leptons and the neutral scalar emerge. The
contribution from these vertices to the collective excitations and
their damping rates is an important future direction of study.

This analysis will require a deeper understanding of the validity of
the perturbative expansion as well as the issue of gauge invariance
as higher order corrections are contemplated.

Finally, going beyond the standard model, it becomes important to
study the effect of these novel phenomena upon neutrino oscillation
and in particular the possibility of MSW resonances. An intriguing
possibility is that of oscillations between different
\emph{branches} as well as the competition between the oscillation
and relaxation time scales. Furthermore small neutrino masses will
lead to helicity flip transitions which may in turn lead to
radiative transitions between the quasiparticle and quasihole
branches (or their antiparticles).

These are currently the focus of our current studies on which we
expect to report in a forthcoming article\cite{proximo}.

\begin{acknowledgments} The author thanks the US NSF for support under
grant PHY-0242134, and Brad Keister,  Adam Leibovich, and Chiu-Man
Ho for illuminating conversations and remarks.
\end{acknowledgments}

\appendix
\section{Real time propagators and self energies.}

\subsection{Fermions}

For a generic massless fermion field $f(\vx,t)$ the Wightmann and
Green's functions at finite temperature are the following

\bea i S^>_{\alpha,\beta} (\vx-\vx',t-t') & = & \langle
f_{\alpha}(\vx,t) \overline{f}_{\beta}(\vx',t')\rangle =
\frac{1}{V}\sum_{\vp}e^{i \vp\cdot(\vx-\vx')}
\;iS^>_{\alpha,\beta}(\vp,t-t')
\label{sgreat}\\
i S^<_{\alpha,\beta} (\vx-\vx',t-t') & = & -\langle
\overline{f}_{\beta}(\vx',t') f_{\alpha}(\vx,t) \rangle =
\frac{1}{V}\sum_{\vp}e^{i \vp\cdot(\vx-\vx')}
\;iS^<_{\alpha,\beta}(\vp,t-t') \label{sless}\eea

\noindent where $\alpha,\beta$ are Dirac indices and $V$ is the
quantization volume.

The real time Green's functions along the forward $(+)$ and backward
$(-)$ branches are given in terms of these Wightmann functions as
follows

\bea \langle f^{(+)}_{\alpha}(\vx,t)
\overline{f}^{(+)}_{\beta}(\vx',t')\rangle & = &
iS^{++}(\vx-\vx',t-t')= i S^>(\vx-\vx',t-t')\Theta(t-t')+ i
S^<(\vx-\vx',t-t')\Theta(t'-t)
\label{Spp} \\
\langle f^{(+)}_{\alpha}(\vx,t)
\overline{f}^{(-)}_{\beta}(\vx',t')\rangle & = &
iS^{+-}(\vx-\vx',t-t')= i S^<(\vx-\vx',t-t')\eea

At finite temperature $T$ it is straightforward to obtain these
correlation functions by expanding the free Fermion fields in terms
of Fock creation and annihilation operators and massless spinors.
The result is conveniently written  in a dispersive form

\bea iS^>_{\alpha,\beta}(\vp,t-t') & = & \int_{-\infty}^{\infty}
dp_0 \rho^>_{\alpha,\beta}(\vp,p_0) e^{-ip_0(t-t')} \label{Sgreatdis}\\
iS^<_{\alpha,\beta}(\vp,t-t') & = & \int_{-\infty}^{\infty} dp_0
\rho^<_{\alpha,\beta}(\vp,p_0) e^{-ip_0(t-t')} \label{Slessdis}\eea

\noindent where

\bea \rho^>_{\alpha,\beta}(\vp,p_0) & = & [1-N_f(p_0)]
\rho^f_{\alpha,\beta}(\vp,p_0) \label{rhogreat}\\
\rho^<_{\alpha,\beta}(\vp,p_0) & = & N_f(p_0)
\rho^f_{\alpha,\beta}(\vp,p_0). \label{rholess}\eea

and $N_f(p_0)$ is the Fermi-Dirac distribution function

\be  N_f(p_0) = \frac{1}{e^{\frac{p_0}{T}}+1} \ee

The free Fermion spectral density is given by

\bea \rho^f(\vp,p_0)  & = &
\frac{\not\!{p}_+}{2p}\delta(p_0-p)+\frac{\not\!{p}_-}{2p}\delta(p_0+p)\label{rhofree}\\
{\not\!{p}_{\pm}} & = & \gamma^0\,p \mp \vec{\gamma}\cdot \vp
\label{ppm} \eea

\subsection{Vector bosons}

Consider a generic real vector boson field $A_{\mu}(\vx,t)$ of mass
$M$. In unitary gauge it can be expanded in terms of Fock creation
and annihilation operators of \emph{physical} states with three
polarizations as

\be A^{\mu}(\vx,t)= \frac{1}{\sqrt{V}}\sum_{\lambda} \sum_{\vk}
\frac{\epsilon^{\mu}_{\lambda}(\vk)}{\sqrt{2\omega_k}}\left[a_{\vk,\lambda}e^{-i\omega_k\,
t}\,e^{i \vk\cdot\vx}+a^{\dagger}_{\vk,\lambda}e^{i\omega_k\,
t}\,e^{-i \vk\cdot\vx}\right] ~~;~~
k^{\mu}\epsilon_{\mu,\lambda}(\vk) =0 \ee

\noindent where $\omega_k=\sqrt{k^2+M^2}$ and $k^{\mu}$ is
\emph{on-shell} $k^{\mu}=(\omega_k,\vk)$. The three polarization
vectors are such that

\be \sum_{\lambda=1}^3 \epsilon^{\mu}_{\lambda}(\vk)
\epsilon^{\nu}_{\lambda}(\vk)= P^{\mu \nu}(\vk)= - \left( g^{\mu
\nu}- \frac{k^\mu k^\nu}{M^2}\right) \label{projector}\ee

It is now straightforward to compute the Wightmann functions of the
vector bosons in a state in which the physical degrees of freedom
are in  thermal equilibrium at temperature $T$. These are given  by

\bea \langle A_{\mu}(\vx,t) A_{\nu}(\vx',t') \rangle & = &
iG^>_{\mu,\nu}(\vx-\vx',t-t') \label{AAgreat} \\
\langle  A_{\nu}(\vx',t') A_{\mu}(\vx,t)\rangle & = &
iG^<_{\mu,\nu}(\vx-\vx',t-t') \label{AAless}\eea

\noindent where $G^{<,>}$ can be conveniently written as spectral
integrals in the form

\bea iG^>_{\mu,\nu}(\vx-\vx',t-t') & = & \frac{1}{{V}}\sum_{\vk}
e^{i\vk\cdot(\vx-\vx')}  \int_{-\infty}^{\infty}dk_0
e^{-ik_0(t-t')}\,[1+N_b(k_0)]\rho_{\mu \nu}(k_0,\vk) \label{Ggreat}\\
iG^<_{\mu,\nu}(\vx-\vx',t-t') & = & \frac{1}{{V}}\sum_{\vk}
e^{i\vk\cdot(\vx-\vx')} \int_{-\infty}^{\infty}dk_0
e^{-ik_0(t-t')}\,N_b(k_0)\rho_{\mu \nu}(k_0,\vk) \label{Gless}\eea

\noindent where

\be N_b(k_0)= \frac{1}{e^{\frac{k_0}{T}}-1}\label{BEdist}\ee

\noindent and the spectral density is given by

\be \rho_{\mu \nu}(k_0,\vk) = \frac{1}{2\omega_k}\left[P_{\mu
\nu}(\vk)\,\delta(k_0-\omega_k)- P_{\mu
\nu}(-\vk)\,\delta(k_0+\omega_k)\right]\label{vbspecdens} \ee

In terms of these Wightmann functions the real time correlation
functions along the forward and backward time branches are given by

\bea \langle A^{(+)}_{\mu}(\vx,t) A^{(+)}_{\nu}(\vx',t') \rangle & =
& iG^>_{\mu,\nu}(\vx-\vx',t-t')\Theta(t-t')+
iG^<_{\mu,\nu}(\vx-\vx',t-t')\Theta(t'-t) \label{App}\\
\langle A^{(+)}_{\mu}(\vx,t) A^{(-)}_{\nu}(\vx',t') \rangle & = &
iG^<_{\mu,\nu}(\vx-\vx',t-t') \label{Apm} \eea

For the charged vector bosons the correlation functions can be found
simply from those of the real vector boson fields described above by
writing the charged fields as linear combinations of \emph{two} real
fields $A^{1,2}$, namely

\be W^{\pm}_{\mu}(\vx,t) = \frac{1}{\sqrt{2}} (A^{1}_{\mu}(\vx,t)\pm
i A^{2}_{\mu}(\vx,t)) \ee

It is straightforward to find the correlation function

\be \langle W^{+}_{\mu}(\vx,t)W^{-}_{\mu}(\vx',t') \rangle =
G^>_{\mu\nu}(\vx-\vx',t-t') \ee

\noindent and similarly for the other necessary Wightmann and
Green's functions.

\subsection{Retarded self-energies}

\begin{figure}[h!]
\begin{center}
\includegraphics[height=3in,width=4in,keepaspectratio=true]{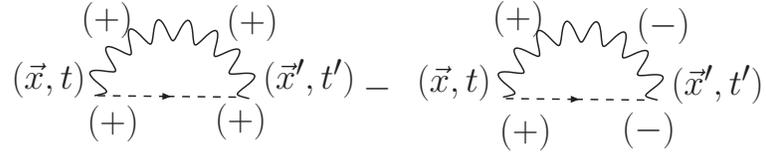}
\caption{Retarded self-energy for charged current interactions. The
wiggly line is a charged vector boson and the dashed line a lepton.
The labels $(\pm)$ correspond to the forward ($+$) and backward
($-$) time branches. The corresponding propagators are
$iS^{\pm,\pm}(\vx-\vx',t-t')$ and $iG^{\pm
\pm}_{\mu\nu}(\vx-\vx',t-t')$ for leptons and charged bosons
respectively. } \label{retSE}
\end{center}
\end{figure}

The diagrams for the one-loop retarded self-energy from charged
current interactions are displayed in fig. (\ref{retSE}). A
straightforward calculation yields for the charged current
contribution the following result

\be \Sigma^{CC}_{ret}(\vx-\vx',t-t') = \frac{ig^2}{2} R\,
\gamma^{\mu}\left[iS^{++}(\vx-\vx',t-t')iG^{++}_{\mu\nu}(\vx-\vx',t-t')-iS^<(\vx-\vx',t-t')iG^{<}_{\mu\nu}(\vx-\vx',t-t')
\right]\gamma^\nu\, L \label{retaSE} \ee

\noindent with

$$R=\frac{(1+\gamma^5)}{2}~~;~~L=\frac{(1-\gamma^5)}{2}.$$

 A similar
result is obtained for the neutral current contribution to the self
energy by simply replacing $g/\sqrt{2} \rightarrow g/2\cos\theta_w$
and $M_W\rightarrow M_Z=M_W/\cos\theta_w $.

Using the representation of the fermion and vector boson propagators
given above the retarded self-energy (\ref{retaSE}) can be written
as

\be \Sigma_{ret}(\vx-\vx',t-t') = \frac{i}{V}\sum_{\vk}
\int_{-\infty}^{\infty} dk_0 \, R
\,\left[\widetilde{\Sigma}_W(\vk,k_0)+\widetilde{\Sigma}_Z(\vk,k_0)\right]
 \, L \, e^{i \vk\cdot(\vx-\vx')}\,e^{-i k_0 (t-t')} \Theta(t-t'), \ee

The contribution from charged  and neutral vector bosons are given
by

\be \widetilde{\Sigma}_W(\vk,k_0) = \frac{g^2}{2} \int \dbarq \int
dp_0 \int dq_0 \,\delta(p_0+q_0-k_0)\,  \gamma^\mu \,
\rho^f(\vk-\vq,p_0) \,\rho^W_{\mu \nu}(\vk)(\vq,q_0)\gamma^\nu\,
(1-N_f(p_0)+N_b(q_0))   \label{retSWdisp}\ee

\be \widetilde{\Sigma}_Z(\vk,k_0) = \frac{g^2}{4\cos^2\theta_w} \int
\dbarq \int dp_0 \int dq_0 \,\delta(p_0+q_0-k_0)\, \gamma^\mu\,
\rho^f(\vk-\vq,p_0)\,\rho^Z_{\mu \nu}(\vq,q_0)\, \gamma^\nu
\,(1-N_f(p_0)+N_b(q_0))   \label{retSZdisp}\ee

\noindent where $\rho_{W,Z}(\vq,q_0)$ are the vector boson spectral
densities given by (\ref{vbspecdens})  with $M\equiv M_{W,Z}(T)$
respectively. It is clear that $\widetilde{\Sigma}_{W,Z}(\vk,k_0))$
correspond  to a vector-like theory.

Using the integral representation of the function $\Theta(t-t')$ the
retarded self energy can be written in the following simple
dispersive form

\be \Sigma_{ret}(\vx-\vx',t-t') =   \frac{1}{V}\sum_{\vk}
\int_{-\infty}^{\infty} \frac{d\omega}{2\pi}\,e^{i
\vk\cdot(\vx-\vx')}\, e^{-i\omega(t-t')} R\,\left[
\Sigma_W(\vk,\omega)+ \Sigma_Z(\vk,\omega) \right]\, L
\label{retSE2}\ee

\be \Sigma_{W,Z}(\vk,\omega) = \int dk_0
\frac{\widetilde{\Sigma}_{W,Z}(\vk,k_0)}{k_0-\omega-i\epsilon}
\label{retSEfin} \ee

\noindent where $\epsilon \rightarrow 0^+$.

Hence from the above expression we identify

\be \widetilde{\Sigma}_{W,Z}(\vk,\omega) = \frac{1}{\pi}
\textit{Im}\,\Sigma_{W,Z}(\vk,\omega)\label{imparts}\ee

Furthermore, since we are considering massless neutrinos and leptons
in the high temperature approximation, the fermionic spectral
density is proportional to $\gamma$ matrices and does not feature
the identity matrix or $\gamma^5$ (see eqns.
(\ref{rhofree}),(\ref{ppm}) therefore there is the following
simplification

\be R\,\left[ \widetilde{\Sigma}_W(\vk,\omega)+
\widetilde{\Sigma}_Z(\vk,\omega) \right]\, L = \left[
\Sigma_W(\vk,\omega)+ \Sigma_Z(\vk,\omega) \right]\, L \label{iden}
\ee

\end{document}